\documentclass[manuscript]{acmart}
\usepackage{subcaption}
\usepackage{xspace}
\usepackage{amsmath,amsfonts}
\usepackage{hyperref}
\usepackage{hyperxmp}
  
\usepackage{amssymb}
\usepackage{graphicx}
\usepackage{textcomp}
\usepackage{xcolor}
\usepackage{booktabs}
\usepackage{multirow}
\usepackage{caption}
\usepackage{dcolumn}
\usepackage{wrapfig}
\definecolor{mygray}{RGB}{247,247,247}
\usepackage{enumitem}
\usepackage{listings}
\usepackage{algorithm}
\usepackage{algpseudocode}
\usepackage{xcolor}
\usepackage[capitalize]{cleveref}
\crefname{section}{Sec.}{Secs.}
\crefname{section}{Section}{Sections}
\crefname{table}{Table}{Tables}
\crefname{table}{Tab.}{Tabs.}
\usepackage{tcolorbox}
\newcommand{\mybox}[1]{%
  \begin{tcolorbox}[colback=mygray,colframe=black,lowerbox=invisible,savelowerto=\jobname_ex.tex]
    \emph{#1}
  \end{tcolorbox}
}
\definecolor{codegreen}{rgb}{0,0.6,0}
\definecolor{codegray}{rgb}{0.5,0.5,0.5}
\definecolor{codepurple}{rgb}{0.58,0,0.82}
\definecolor{backcolour}{rgb}{0.95,0.95,0.92}
\lstdefinestyle{mystyle}{
  backgroundcolor=\color{backcolour},   commentstyle=\color{codegreen},
  keywordstyle=\color{magenta},
  numberstyle=\tiny\color{codegray},
  stringstyle=\color{codepurple},
  basicstyle=\ttfamily\footnotesize,
  breakatwhitespace=false,         
  breaklines=true,                 
  captionpos=b,                    
  keepspaces=true,                 
  numbers=left,                    
  numbersep=5pt,                  
  showspaces=false,                
  showstringspaces=false,
  showtabs=false,                  
  tabsize=2
}

\setcopyright{cc}
\setcctype{by}
\acmJournal{TOSEM}
\acmYear{2026} \acmVolume{1} \acmNumber{1} \acmArticle{}
\acmMonth{1} \acmDOI{10.1145/3805043}

\begin{document}

\title[Pareto-Gated Credit Assignment for Concise Unit-Test Generation]{Ockhamareto: Pareto-Gated Segment-Level Credit Assignment for Concise Unit-Test Generation with Reinforcement Learning}

\author{Dong HUANG}
\email{dhuang@nus.edu.sg}
\affiliation{%
  \institution{National University of Singapore}
  \country{Singapore}
}
\author{Mark Harman}
\email{mark.harman@ucl.ac.uk}
\affiliation{%
  \institution{University College London}
  \country{London, UK}
}
\author{Jie M. Zhang}
\email{jie.zhang@kcl.ac.uk}
\affiliation{%
  \institution{King's College London}
  \country{London, UK}
}

\author{Zhijiang Guo}
\email{zhijiangguo@hkust-gz.edu.cn}
\affiliation{%
  \institution{Hong Kong University of Science and Technology (Guangzhou)}
  \country{Guangzhou, CN}
}
\author{Mingzhe Du}
\authornote{Corresponding author.}
\email{mingzhe@nus.edu.sg}
\affiliation{%
  \institution{National University of Singapore}
  \country{Singapore}
}
\author{See-Kiong Ng}
\email{seekiong@nus.edu.sg}
\affiliation{%
  \institution{National University of Singapore}
  \country{Singapore}
}

\begin{abstract}
% LLM-generated unit tests reach high coverage but tend to be \emph{bloated}:
% dozens of near-duplicate tests that inflate review, CI, and maintenance cost
% while catching few real bugs. Prior work treats bug-catching power (mutation
% score) and suite size (test count) as \emph{competing} objectives to be traded
% off: catch more bugs by writing more tests, or shrink the suite and settle for
% less. We show that with the right reward shape, the two are jointly
% achievable: a single policy can simultaneously raise mutation \emph{and} shrink
% suites.

We introduce \textbf{Ockhamareto}, a single-shot GRPO framework for unit-test generation and selection, based on the principles of \emph{Ockham's Razor} and \emph{Pareto Optimality}. 
Ockhamareto has two principal components: 
(i)~a \emph{Pareto-gated Bonus} that rewards only rollouts non-dominated in~(mutation, $-$\#tests) space, and (ii)~\emph{Token-level Segment Credit}, which attributes each test's marginal mutation kills back to the tokens of its unit-test block. 
On the \emph{UnLeakedTestBench~(ULT)}, Ockhamareto \emph{strictly Pareto-dominates} the strongest RL baseline~(\emph{MIST-RL}).
Furthermore, it dominates on {\em each and all} optimization objectives, 
catching more bugs ($49.9\%$ vs $31.3\%$ mutation score at $N{=}5$), using
\emph{fewer} tests ($2.60$ vs $4.67$ on average), thereby achieving  $3.4\times$ the per-test trade-off improvement.
The advantage is found in all four benchmarks~(\emph{HumanEval+}, \emph{MBPP+}, \emph{CodeContests}, \emph{TestGenEval-Lite}): 
Ockhamareto leads both mutation and coverage metrics on every one, always with the smallest suite. 
Ockhamareto also outperforms the state-of-the-art at all  model scales, adding
$+30$--$35$~pp mutation at 4B, 9B, and 27B model sizes.
We also show that the knee point of the optimal trade-off between efficiency and effectiveness on the Pareto front is not correlated with obvious more easily computed proxy metrics, such as function size.
This finding motivates the Pareto front computation; 
it is needed to identify this crucial engineering trade-off for each function under test.
% and Ockhamareto-4B already
% beats \emph{base}-27B ($30.5\%$) with $\approx 7\times$ fewer parameters and
% $28\%$ smaller suites. 
% Reward \emph{shape}, not scale, is the primary lever;
% the two compose.
\end{abstract}




\maketitle

\section{Introduction}
\label{sec:intro}

Software testing inevitably faces diminishing returns. Additional tests may
reveal additional faults, but they also increase execution, review, and
maintenance cost. The practical objective is therefore not simply to generate
as many tests as possible, but to obtain test suites that achieve strong
fault-detection effectiveness without unnecessary redundancy. This
effectiveness--effort trade-off is a longstanding problem in software testing;
we revisit its Pareto formulation and the associated question of diminishing
returns in \S\ref{sec:bg_pareto}.

The emergence of large language models (LLMs) makes this problem particularly
timely. LLMs can generate readable and idiomatic unit tests directly from
source code and documentation. Early approaches primarily rely on prompt
engineering and iterative refinement. ChatTester first elicits the intention of
the focal method and repairs uncompilable tests
\citep{yuan2024chattester}; SymPrompt issues coverage-guided prompts for
different static execution paths \citep{ryan2024symprompt}; CodaMosa uses LLMs
to escape coverage plateaus in search-based test generation
\citep{lemieux2023codamosa}; and TestPilot mines usage examples to guide
generation and re-generation \citep{schafer2024testpilot}. These methods
substantially improve the ability of LLMs to produce executable and
coverage-reaching tests.

Prompt-based generation, however, optimizes test quality only indirectly: the
model itself is not trained against the execution outcome that ultimately
matters. This limitation has motivated reinforcement learning (RL) from
execution feedback, in which generated tests are executed and their observed
behavior provides the reward signal. Recent work rewards structural coverage
\citep{testctrl2025}, marginal coverage gain
\citep{testdecision2026}, or marginal mutation kills
\citep{zhu2026mist}. Mutation score is particularly attractive because it
measures whether a test suite distinguishes a reference implementation from
faulty variants and has been shown to correlate with real-fault detection more
strongly than structural coverage alone
\citep{just2014mutants,jia2011mutation,papadakis2019mutation}.

Yet an important efficiency problem remains. The strongest recent RL
approaches obtain fine-grained per-test feedback through \emph{multi-turn}
generation: the model emits one test, receives execution feedback, and then
generates another. This makes per-test credit assignment straightforward, but
has two consequences. First, generating an $n$-test suite requires multiple
sequential LLM interactions. Second, the policy is trained to determine whether
each newly generated test adds value, rather than whether the \emph{suite as a
whole} provides a good effectiveness--size trade-off. As long as another test
produces some marginal gain, generation can continue. The resulting suites can
therefore contain many tests whose additional fault-detection value is small
relative to their review, execution, and maintenance burden.

A natural alternative is \emph{single-shot} generation, in which one model call
produces the complete test suite. Single-shot generation avoids sequential
inference and requires the policy to decide which tests are sufficiently
valuable to include in the suite. It also creates a harder learning problem.
Standard Group Relative Policy Optimization (GRPO)
\citep{shao2024deepseekmath} assigns a scalar advantage to an entire generated
trajectory. When that trajectory contains several independently valuable test
functions, the policy learns only that ``this suite was good''; it cannot tell
which individual tests contributed to fault detection and which were
redundant.

This exposes two coupled optimization problems at different granularities.
At the \emph{suite level}, the learner must distinguish suites that provide a
good effectiveness--size trade-off from suites that achieve similar
effectiveness using unnecessary tests. At the \emph{test level}, it must
identify which tests inside a generated suite are responsible for its
fault-detection capability. A single scalar trajectory reward provides neither
an explicit whole-suite conciseness criterion nor fine-grained attribution
among the tests within that trajectory.

We address these problems with \textbf{Ockhamareto}, a single-shot GRPO
framework for concise and effective unit-test generation. Ockhamareto combines
two complementary mechanisms.
First, at the \emph{suite level}, we introduce a
\textbf{Pareto-gated group-relative bonus}. Within each GRPO group, a rollout
receives the conciseness bonus only when it is non-dominated with respect to
test effectiveness and suite size. A suite that uses more tests without
providing greater effectiveness is therefore not rewarded for conciseness,
whereas suites representing genuinely different effectiveness--size
trade-offs remain eligible. Unlike a fixed penalty on the number of tests, the
Pareto gate does not require committing to a universal exchange rate between
one additional test and one additional unit of fault detection.
Second, at the \emph{test level}, we introduce
\textbf{token-level segment credit}. We attribute each test's marginal mutation
kills to the token span corresponding to that test and inject this information
as a per-token advantage offset on top of the suite-level GRPO signal.
High-yield tests receive positive intra-trajectory credit, whereas redundant
or failing tests receive lower or negative credit. Segment credit therefore
gives a single-shot policy the fine-grained per-test supervision that
multi-turn methods obtain structurally, without requiring one sequential LLM
interaction per test. In short, the Pareto gate asks whether the
\emph{whole suite} is worth its size, while segment credit identifies
\emph{which tests within the suite} create its value.

Our empirical results show that these two mechanisms can improve effectiveness
and conciseness simultaneously. On UnLeakedTestBench (ULT),
Ockhamareto-4B achieves $49.9\%$ mutation score with an average of $2.60$
tests at $N{=}5$, compared with $31.3\%$ mutation and $4.67$ tests for the
strongest RL baseline, MIST-RL. Ockhamareto therefore improves mutation by
$18.6$ percentage points while using $44\%$ fewer tests. Its mutation score
largely saturates within the first three tests, and its first test alone
achieves $33.3\%$ mutation, already exceeding MIST-RL's $31.3\%$ mutation at
five tests. This front-loading behavior is consistent with the intended effect
of segment credit: concentrating learning signal on tests that contribute the
largest marginal fault-detection value.

The advantage generalizes beyond ULT. Across HumanEval+, MBPP+,
CodeContests, and TestGenEval-Lite, Ockhamareto achieves the highest mutation
score and statement/branch coverage among the evaluated methods while using
the smallest suites. The same reward design also transfers across model
scales: at 4B, 9B, and 27B parameters, Ockhamareto improves mutation over the
corresponding untuned model by approximately $30$--$35$ percentage points.
Notably, Ockhamareto-4B reaches $49.9\%$ mutation compared with $30.5\%$ for
the untuned 27B model, showing that model scale alone does not replace the
benefit of reward design.

Beyond improving test generation itself, the Pareto formulation also lets us
revisit a longstanding engineering question: \emph{how many tests are worth
maintaining for an individual function?} We construct empirical
(mutation, \#tests) Pareto fronts from generated suites and study their
\emph{knee points}, beyond which additional tests yield diminishing
fault-detection returns. On a random 100-function subset of ULT, the median
knee occurs at three tests, but individual knees range from one to fourteen
tests. Moreover, knee location is not significantly associated with simple
static proxies such as lines of code or cyclomatic complexity. This suggests
that a fixed rule such as ``larger functions require more tests'' cannot
reliably determine an appropriate suite size; the effectiveness--size trade-off
must instead be established empirically for the function under test.
Ockhamareto does not itself define a universally optimal stopping point.
Rather, it generates substantially stronger low-cost candidate suites and
contributes the majority of solutions on the pooled empirical Pareto fronts,
making the function-specific trade-off easier to expose and navigate.
Our contributions are:

\begin{itemize}[leftmargin=1.4em,itemsep=2pt,topsep=2pt]

    \item \textbf{Joint effectiveness--conciseness formulation.}
    We formulate single-shot unit-test generation as a joint optimization
    problem over fault-detection effectiveness and suite size, and show
    empirically that better fault detection need not require larger generated
    suites. Ockhamareto shifts the observed
    (mutation, $-$\#tests) frontier outward relative to existing RL baselines.

    \item \textbf{Pareto-gated suite-level optimization.}
    We introduce a group-relative Pareto gate that awards a conciseness bonus
    only to non-dominated rollouts, allowing GRPO to optimize whole-suite
    effectiveness and size without specifying a fixed scalar exchange rate
    between the two objectives.

    \item \textbf{Token-level segment credit.}
    We introduce fine-grained credit assignment for single-shot test-suite
    generation by mapping each test's marginal mutation kills onto its own
    token span. This provides intra-trajectory per-test supervision while
    retaining single-shot inference.

    \item \textbf{Comprehensive empirical evaluation and Pareto-front analysis.}
    Across five held-out benchmarks and three model scales, we show that
    Ockhamareto consistently improves fault detection while producing smaller
    suites. We further characterize per-function empirical Pareto fronts and
    show that their knee locations vary substantially and are not reliably
    predicted by simple static code metrics.

\end{itemize}

\section{Background}
\label{sec:background}

This section establishes the conceptual and technical foundations of our
approach. We first formulate the longstanding effectiveness--effort problem in
software testing through Pareto optimality and diminishing returns. We then
review mutation testing, reinforcement learning for code generation, and the
credit-assignment problem that arises when an entire test suite is generated
as a single trajectory.

\subsection{Testing Effectiveness, Effort, and Pareto Optimality}
\label{sec:bg_pareto}

A fundamental practical question in software testing is:

\begin{quote}
``When can I stop testing?''
\end{quote}

The question cannot generally be answered by establishing that all faults have
been excluded. Exhaustive testing is impractical for most non-trivial
programs, and successful execution on a finite set of inputs does not imply
correct behavior on every untested input. This limitation is captured by
Dijkstra's well-known observation that testing can demonstrate the presence of
bugs, but not their absence \cite{dijksra:aphorism}. Questions about the
limits of testing and program checking have accompanied the discipline since
its early development \cite{turing:checking}.

For engineering purposes, however, the absence of a definitive scientific
stopping criterion does not make the decision meaningless. Instead, it
transforms the question into one of diminishing returns:

\begin{quote}
``At what point does the fault-detection benefit of additional testing cease to justify its cost?''
\end{quote}

This formulation separates two quantities. \emph{Effectiveness} captures the
fault-revealing capability of a test suite, while \emph{effort} captures the
resources required to construct, execute, review, and maintain it. Effort can
be represented in many ways, including execution time, monetary cost,
maintenance burden, or developer attention. In this work, we use the
\emph{number of tests} as a simple and directly observable proxy for suite
cost. This proxy does not capture every component of testing effort, but it
directly reflects the redundancy problem that arises when generative models
produce increasingly large suites.

When effectiveness and effort are considered jointly, there need not be a
single solution that is best on both dimensions. Instead, the natural object
is a \emph{Pareto front}. Given two candidate suites, suite $A$ dominates
suite $B$ if $A$ is at least as effective while using no more tests, and is
strictly better on at least one of these dimensions. A suite is
\emph{Pareto-optimal} if no available alternative dominates it
\citep{Pareto1906}. The resulting Pareto front retains only suites for which
improving one objective requires sacrificing the other.

Pareto reasoning has previously been applied to software testing. In
particular, Yoo and Harman used Pareto optimality to guide test selection from
an existing test suite \citep{symh:issta07}. Their setting assumes that a
collection of tests already exists and asks which subset should be retained
under competing objectives. Our setting differs in an important respect:
generation and selection are no longer cleanly separated. An LLM policy
constructs the suite itself, so the training objective can influence both
\emph{which tests are generated} and \emph{how many tests are generated}.
This makes it possible to move conciseness from a post-hoc selection criterion
into the learning process itself.

A Pareto front also exposes the notion of a \emph{knee point}. Moving along
the front initially may yield large improvements in effectiveness for a small
increase in effort, whereas beyond some region additional tests provide only
small gains. The knee characterizes this transition into diminishing returns.
It therefore provides a useful engineering operating point, although it should
not be interpreted as a universal or mathematically unique definition of when
testing must stop.

This view connects naturally to the principle commonly associated with
Ockham's razor \citep{Ockham1323}: additional complexity should be retained
only when it provides sufficient explanatory or practical value. Applied to
testing, the corresponding objective is not simply to minimize the number of
tests. An extremely small suite with poor fault detection is not desirable.
Rather, the goal is to avoid tests whose additional cost is not justified by
additional effectiveness---as few tests as possible, but no fewer than are
needed to preserve valuable fault detection.

These ideas motivate two distinct uses of Pareto analysis in our work.
During \emph{training}, Ockhamareto uses Pareto dominance within each GRPO
group as a reward gate, encouraging policies that generate effective but
concise suites. During \emph{analysis}, we construct empirical per-function
Pareto fronts from sampled suites to examine the available
effectiveness--size trade-offs and their knee points. The former is a learning
mechanism; the latter is an engineering analysis of the solutions produced by
the learned policies. Keeping these two roles distinct is important:
Ockhamareto improves the candidate suites available near the frontier, while
the empirical front provides the information needed to choose among different
operating points.

\subsection{Unit Testing and Mutation Testing}

Unit testing validates individual functions or methods in isolation. A unit
test suite consists of test cases that exercise a target function with
specific inputs and assert expected behavior. Two widely used criteria for
assessing such suites are \emph{structural coverage}, including statement and
branch coverage, and \emph{mutation score}.

Structural coverage measures the proportion of program elements executed by a
test suite. Although coverage is inexpensive to compute and widely used, it
does not directly establish that the executed behavior has been meaningfully
checked. For example, an assertion such as
\texttt{f(x) is not None} may execute substantial portions of a function while
providing little discrimination between correct and faulty behavior.

Mutation testing \citep{jia2011mutation,papadakis2019mutation} evaluates a
suite by introducing small syntactic changes, or \emph{mutants}, into the
program under test. A mutant is \emph{killed} when at least one test detects
the behavioral difference and fails on the mutated implementation. The
\emph{mutation score} is the fraction of mutants killed by the suite.
Empirical studies have shown a strong relationship between mutation score and
real-fault detection \citep{just2014mutants}, making mutation testing a useful
fault-oriented adequacy criterion despite its higher computational cost.

For test generation, mutation testing can play two roles. It can serve as an
evaluation metric for comparing generated suites, and it can provide an
execution-derived reward signal during learning. Ockhamareto uses mutation
information in both roles. At the suite level, mutation contributes to the
quality signal used during GRPO training; at the test level, marginal mutation
kills provide the fine-grained signal used for segment credit.

\subsection{Reinforcement Learning for Code Generation}

Reinforcement learning from execution feedback treats code generation as a
sequential decision problem. A language model generates code token by token,
the generated artifact is executed or otherwise verified, and the resulting
outcome is converted into a reward used to update the policy. Verifiable
execution signals are particularly attractive for code because they provide
objective feedback without requiring a learned reward model.

Group Relative Policy Optimization (GRPO)
\citep{shao2024deepseekmath} is a policy-optimization method suited to such
verifiable-reward settings. Instead of training a separate value model, GRPO
samples a group of $K$ candidate rollouts for the same task and computes the
advantage of each rollout relative to other members of the group. For rewards
$\{r_j\}_{j=1}^{K}$, a simplified group-relative advantage can be written as

\begin{equation}
A_i = r_i - \frac{1}{K}\sum_{j=1}^{K} r_j .
\end{equation}

The policy is then updated to increase the likelihood of trajectories with
positive relative advantage and decrease the likelihood of those with
negative relative advantage, subject to regularization against a reference
policy.
This group-relative structure is useful for our problem because test suites
generated for the same function can be compared directly. It also provides a
natural setting for Pareto gating: rather than defining a global threshold for
an acceptable suite, Ockhamareto determines whether a rollout is dominated by
other candidates generated for the same task.

Standard trajectory-level GRPO nevertheless assigns the same scalar advantage
to all action tokens within one rollout. This assumption becomes problematic
when a trajectory is internally structured and its components make very
different contributions to the observed reward. A generated test suite is
precisely such a trajectory.

\subsection{The Credit-Assignment Challenge in Test-Suite Generation}
\label{sec:bg_credit}

A generated test suite is an internally structured trajectory: individual
tests may kill many previously undetected mutants, duplicate behavior already
covered by earlier tests, or fail on the reference implementation. A single
trajectory reward collapses these distinct contributions. Multi-turn methods
make per-test attribution straightforward by executing tests sequentially, but
require repeated model interactions; single-shot generation retains one-call
inference while losing that structural source of credit assignment.

The suite structure lets us recover the missing information directly. Each
unit test block can be executed and assigned a marginal
fault-detection contribution, so Ockhamareto treats each test as a segment and
maps its execution-derived mutation contribution back to the tokens that
created it. The suite-level reward and Pareto gate judge the complete rollout,
whereas segment credit differentiates the tests \emph{within} that rollout.
The next section describes how the two signals are combined.

\section{Method}
\label{sec:method}

\begin{figure*}[t]
\centering
\includegraphics[width=\textwidth]{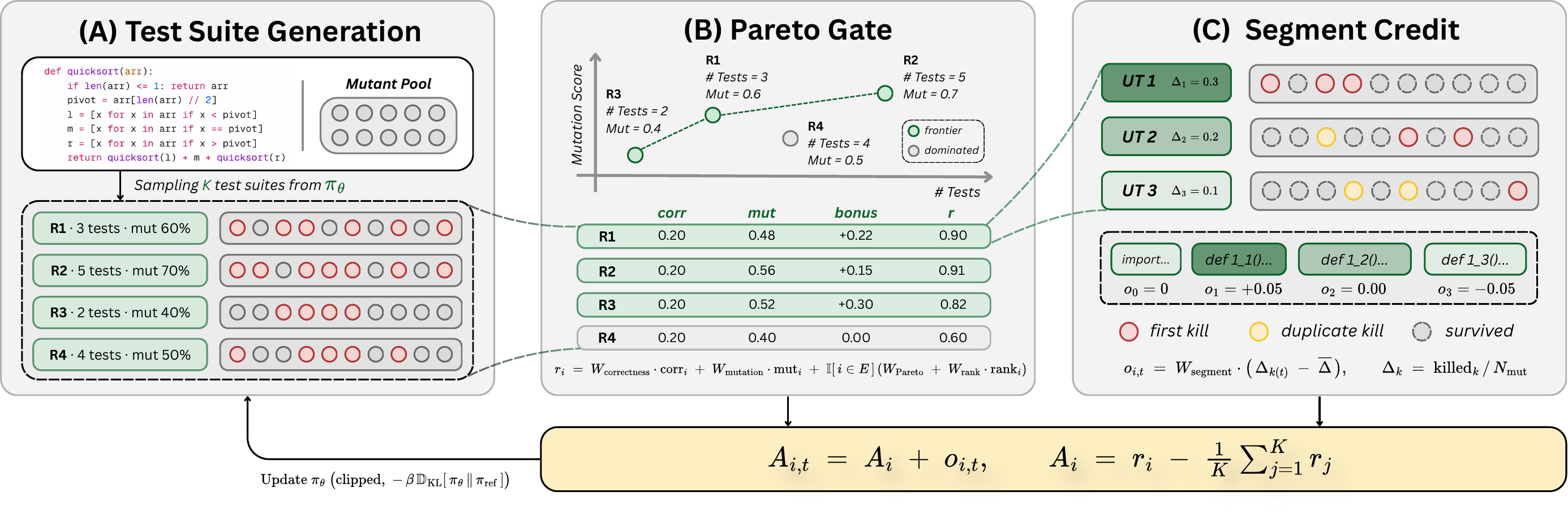}
\caption{Overview of \textbf{Ockhamareto}. \textbf{(A)}~For each task the policy
$\pi_\theta$ samples a GRPO group of $K$ complete \texttt{pytest} suites in a
single shot, and each suite is executed in the sandbox against a prebuilt
mutant pool. \textbf{(B)}~\emph{Pareto gate} (\S\ref{sec:pareto}): only
rollouts non-dominated in (mutation, $-$\#tests) space receive the conciseness
bonus, so $R_4$, which catches fewer mutants than $R_1$ with more tests, is
dominated and earns nothing, while $R_1$--$R_3$ are on the frontier.
\textbf{(C)}~\emph{Segment credit} (\S\ref{sec:segment}): the mutants each test
is the \emph{first} to kill give a per-test score $\Delta_k$, which is
zero-meaned across the suite and mapped onto that test's own token span via the
tokenizer's offset mapping. The update uses per-token advantages
$A_{i,t}=A_i+o_{i,t}$ under a KL constraint to $\pi_{\text{ref}}$, so
high-yield tests are reinforced and redundant ones penalized \emph{within} a
single trajectory.}
\label{fig:overview}
\end{figure*}

\subsection{Problem setup}
Given a target function (its name, signature, docstring, and reference
implementation), the policy must emit a complete \texttt{pytest} suite in
\emph{one} shot. For each task we sample a GRPO group of $K{=}8$ suites and
score all $K$ in the sandbox (Figure~\ref{fig:overview}). The single-shot setting is a deliberate design
choice: it forces the policy to front-load its most discriminating tests
(there is no second chance) and keeps inference cost constant regardless of
suite size.

\subsection{Suite-level quality reward}
\label{sec:quality}
For a rollout $i$ whose suite parses and executes, we define
\begin{equation}
q_i = W_{\text{c}}\cdot \text{corr}_i + W_{\text{m}}\cdot \text{mut}_i ,
\end{equation}
with $W_{\text{c}}{=}0.2$, $W_{\text{m}}{=}0.8$\footnote{Default setting used by existing works.}, and $q_i{=}0$ if the suite
fails to run; $\text{corr}_i$ is the fraction of tests passing on the
reference implementation and $\text{mut}_i$ is the mutation score against a
prebuilt mutant pool. We put \emph{no} weight on coverage: it is highly
correlated with mutation but strictly weaker, and rewarding it invites gaming
via assertions like \texttt{f(x) is not None} that execute a line without
checking semantics.

\subsection{Pareto-gated conciseness bonus}
\label{sec:pareto}
The quality reward says nothing about suite size, so we add a group-relative
bonus that rewards conciseness directly. Within a group, rollout $i$ is
\emph{dominated} by rollout $j$ iff
\begin{equation}
q_j \ge q_i \;\wedge\; n_j \le n_i \;\wedge\; (q_j > q_i \vee n_j < n_i),
\end{equation}
where $n_i$ is the test count. Rollout $i$ is \emph{Pareto-eligible} iff no
other rollout dominates it.
Only eligible rollouts receive the bonus
\begin{equation}
b_i = W_{\text{P}} + W_{\text{N}}\cdot \mathrm{rank}_i,
\end{equation}
where $W_{\text{P}}$ is a flat membership bonus (rewards being on the
frontier at all, including the high-quality/high-$n$ corner) and the
optional rank term $\mathrm{rank}_i\in[0,1]$ linearly favors fewer tests
\emph{within} the frontier. Decoupling membership
from rank lets us ablate the two effects independently. Any suite that
trades quality for size is, by construction, dominated by its
larger-but-better neighbor and earns nothing. 

\subsection{Token-level segment credit}
\label{sec:segment}
GRPO assigns a single scalar advantage to a whole trajectory, so every token in
the suite is credited identically, and the model cannot tell \emph{which} test
did the bug-catching work. But a test suite decomposes naturally: it is a
sequence of independently valuable \texttt{def test\_*} blocks whose per-test
mutation outcomes we already compute in the sandbox. We exploit that structure
by giving GRPO sub-trajectory resolution: on top of the scalar advantage, we
add a per-token offset derived from each test's marginal kills.

Let the sandbox report, for each test $k$, the number of mutants it was the
\emph{first} to kill, $\text{killed}_k$, out of $N_{\text{mut}}$ scored mutants.
We define the per-segment reward
\begin{equation}
\Delta_k =
\begin{cases}
\text{killed}_k / N_{\text{mut}} & \text{if test } k \text{ passes,}\\
-\,\text{SEG\_FAIL\_PENALTY} & \text{if test } k \text{ fails,}\\
0 & \text{otherwise,}
\end{cases}
\end{equation}
We then zero-mean across the segments of
the trajectory, $\tilde\Delta_k = \Delta_k - \overline{\Delta}$, which preserves
GRPO's centering property (the mean offset is $\approx 0$, so the
trajectory-level advantage is unchanged). Finally we distribute
$W_{\text{seg}}\cdot\tilde\Delta_k$ to every token whose character midpoint lies
inside test $k$'s source span:
\begin{equation}
\label{eq:offset}
o_t = W_{\text{seg}}\cdot \tilde\Delta_{k(t)},
\end{equation}
and $o_t{=}0$ for boilerplate, prefixes, or tokens of failed tests' siblings.

To map blocks to tokens robustly, we decode the sampled action, re-tokenize it
\emph{with offset mapping}, and confirm the re-tokenized ids match the sampled
ids; on any mismatch we fall back to the scalar GRPO advantage. We locate
\texttt{def test\_*} spans by regular expression and pair each test with its
character range up to the next test (Algorithm~\ref{alg:segment}, which also gives a worked numerical
example). Because the
sandbox renames tests to \texttt{test\_\textless n\textgreater\_*} in source
order, we match the $k$-th block to sandbox index $k$ by \emph{order}, not name.

Because the centered offsets sum to zero, they leave the trajectory-level GRPO
advantage untouched but reshape the intra-trajectory signal: high-yield tests
receive positive credit and low-yield tests receive negative credit, all inside
a single rollout. The model learns which \emph{kind} of test to write next
time, rather than merely ``write more suites like this one''.
Nothing in the mechanism is specific to \texttt{pytest}, or indeed to test
generation: it applies to any output that parses into independently
scorable segments, given a per-segment score and a span locator.
\texttt{def test\_*} blocks are simply the natural segmentation here;
assertion lines, input--output cases, or reasoning steps would serve the
same role under a different parser.

\paragraph{Putting it together.}
Algorithm~\ref{alg:loop} summarizes one GRPO step: sample a group, score each
suite in the sandbox, add the Pareto-gated group bonus, compute the per-token
segment offsets, and apply the centered, KL-constrained update.

\begin{algorithm}[t]
\caption{One Ockhamareto training step. Lines 6--8 are the Pareto gate
(\S\ref{sec:pareto}); line~11 is segment credit (\S\ref{sec:segment}).}
\label{alg:loop}
\begin{algorithmic}[1]
\Require task $x$, policy $\pi_\theta$, reference $\pi_{\text{ref}}$, group size $K$
\State Sample $K$ suites $\{y_i\}_{i=1}^{K} \sim \pi_\theta(\cdot \mid x)$
\For{$i = 1, \dots, K$}
  \State Sandbox: $\text{corr}_i, \text{mut}_i, n_i$, per-test kills
  \State $q_i \gets W_{\text{c}}\,\text{corr}_i + W_{\text{m}}\,\text{mut}_i$
         \Comment{$0$ if the suite fails to run}
\EndFor
\State $E \gets$ Pareto-eligible set in $(\text{mut}, -n)$ space
\For{$i \in E$}
  \State $q_i \gets q_i + W_{\text{P}} + W_{\text{N}}\,\mathrm{rank}_i$
         \Comment{conciseness bonus}
\EndFor
\State $A_i \gets q_i - \mathrm{mean}_j(q_j)$
       \Comment{GRPO advantage}
\State $A_{i,t} \gets A_i + o_{i,t}$ with segment offsets $o_{i,t}$
       from Eq.~\ref{eq:offset}
\State Update $\theta$ with per-token advantages $A_{i,t}$ under a KL
       constraint to $\pi_{\text{ref}}$
\end{algorithmic}
\end{algorithm}

% \subsection{Training and inference}
% We train with GRPO on a hosted RL service \citep{tinker2025} using LoRA
% \citep{hu2022lora} (rank $32$) on a Qwen3.5-4B base \citep{qwen3}, with a batch
% of $32$ tasks, group size $K{=}8$, and KL-constrained updates; complete
% hyper-parameters are in \cref{sec:hparams}. At inference we sample one
% suite per prompt greedily ($\text{temperature}{=}0.0$, up to $4{,}096$ new
% tokens); a \emph{few-shot} variant prepends the same single worked
% example used to construct the training prompts, which improves format
% adherence at no compute cost. Rewards and evaluation both run in a
% sandboxed harness extended to attribute mutation kills at per-test
% granularity (\cref{sec:infra}).

\subsection{Segment-Credit Details}
\label{sec:segdetails}

Algorithm~\ref{alg:segment} gives the per-token offset
computation of \S\ref{sec:segment} in full, including the
re-tokenization check and the scalar-GRPO fallback.

\begin{algorithm}[t]
\caption{Per-token segment offsets: per-test mutation outcomes are mapped
to source-span tokens via the tokenizer's offset mapping and added on top
of GRPO's scalar advantage.}
\label{alg:segment}
\begin{algorithmic}[1]
\Require sampled token ids $y$, tokenizer, sandbox result (per-test
         first-kills, pass/fail)
\State $\text{text} \gets \mathrm{decode}(y)$;\quad
       $\text{spans} \gets$ char ranges of \texttt{def test\_*} blocks
\State Re-tokenize $\text{text}$ with offset mapping
\If{re-tokenized ids $\neq y$}
  \State \Return $o \gets \mathbf{0}$ \Comment{fall back to scalar GRPO}
\EndIf
\For{each test $k$}
  \State $\Delta_k \gets$ per-segment reward (kills, pass/fail)
\EndFor
\State $\tilde\Delta_k \gets \Delta_k - \overline{\Delta}$
       \Comment{zero-mean, preserves GRPO centering}
\For{each token $t$}
  \State $o_t \gets W_{\text{seg}} \tilde\Delta_{k(t)}$ if $t$'s char
         midpoint lies in span $k(t)$, else $0$
\EndFor
\State \Return $o$
\end{algorithmic}
\end{algorithm}

\paragraph{Worked example.}
Consider a group rollout whose suite has three tests with first-kill counts
$(\text{killed}_1,\text{killed}_2,\text{killed}_3)=(40,2,0)$ out of
$N_{\text{mut}}{=}100$, all passing. Then $\Delta=(0.40,0.02,0.0)$,
$\overline{\Delta}=0.14$, and the centered offsets are
$\tilde\Delta=(+0.26,-0.12,-0.14)$. With $W_{\text{seg}}{=}0.5$, every token of
test~1 receives $+0.13$ on top of the scalar advantage, while tokens of the two
redundant tests receive $-0.06$ and $-0.07$. The policy is thus pushed toward
the high-yield test and away from the redundant ones \emph{within the same
trajectory}, a distinction the scalar GRPO advantage cannot make.

\section{Experimental Setup}
\label{sec:setup}

\subsection{Datasets}
\label{sec:datasets}

\paragraph{Training data.}
We train on a subset of the Possibly-Leaked TestBench (PLT) \citep{huang2026benchmark}, comprising approximately 10{,}500 Python library functions drawn from real-world repositories in The Stack v2. Each function in PLT is a self-contained unit with its signature, docstring, and reference implementation, along with a prebuilt mutant pool generated by \texttt{cosmic-ray} \citep{cosmicray}. We apply strict decontamination: any function whose test cases appear verbatim in common LLM training corpora is excluded from our training set to prevent data leakage. The training set is further filtered to exclude all tasks that appear in any of our five evaluation benchmarks, ensuring a clean held-out evaluation.
The mutant pool for each task contains syntactic variants of the reference
implementation, including operator replacements (e.g., \texttt{+} $\to$
\texttt{-}), constant mutations (e.g., \texttt{0} $\to$ \texttt{1}), and
control-flow alterations (e.g., \texttt{break} $\to$ \texttt{continue}). The
pool is fixed per task and reused across rollouts, providing a reproducible
mutation-scoring environment.

\paragraph{Evaluation benchmarks.}
We evaluate on five held-out benchmarks spanning four distinct code distributions, ensuring that our method generalizes beyond the training domain.

\textbf{(i) UnLeakedTestBench (ULT).} This is our in-distribution evaluation set, also drawn from The Stack v2 but strictly disjoint from the training set. ULT focuses on Python library functions that have been verified to have no test-case leakage in LLM training data. For evaluation, we use a 2{,}126-task subset that excludes tasks whose \texttt{cosmic-ray} mutant pool exceeds our per-rollout time budget (tasks with $>$500 mutants are excluded to keep evaluation tractable). This subset maintains statistical representativeness while ensuring absolute mutation scores remain comparable across methods.

\textbf{(ii) HumanEval+} and \textbf{(iii) MBPP+} \citep{liu2023evalplus}. These are standard function-level Python benchmarks widely used in code-generation research. HumanEval+ contains 164 programming problems with hand-written reference solutions, while MBPP+ contains 399 problems. Both benchmarks were originally designed for code generation evaluation but have been adopted for test generation by recent RL work \citep{zhu2026mist}. We build per-task \texttt{cosmic-ray} mutant pools over the reference implementations to enable mutation-score evaluation.

\textbf{(iv) CodeContests} \citep{li2022codecontests}. This benchmark consists of competitive programming problems drawn from Codeforces, AtCoder, and other contest platforms. Unlike the function-level benchmarks above, CodeContests problems are algorithmic challenges with complex logic and edge cases. Critically, each task in CodeContests comes with a large corpus of \emph{incorrect Python submissions} from human contestants. We use these wrong submissions as the fault source in place of synthetic mutants: a test suite's quality is measured by how many of these real buggy implementations it rejects. This provides a complementary view of fault-detection capability grounded in actual human errors rather than syntactic mutations.

\textbf{(v) TestGenEval-Lite} \citep{jain2024testgeneval}. This is a repo-level benchmark comprising Python test generation tasks extracted from real open-source projects, including Django, sympy, scikit-learn, and matplotlib. Tasks are drawn from the SWE-Bench dataset \citep{jimenez2024swebench} and adapted for test generation. Unlike the function-level benchmarks, TestGenEval-Lite requires reasoning about module-level context and cross-function dependencies. We build \texttt{cosmic-ray} mutant pools over the target functions. Due to the higher complexity and varied execution environments, we report results on the \emph{shared usable subset} where every method produces at least one passing test on the reference module, ensuring fair comparison.

For benchmarks (i), (ii), (iii), and (v), we build per-task \texttt{cosmic-ray} mutant pools; for (iv), the wrong-submission set replaces synthetic mutants entirely. All models are evaluated single-shot with the same prompt template, and the same per-test attribution harness scores every rollout.

\subsection{Models and Training Infrastructure}
\label{sec:models}

\paragraph{Base model and architecture.}
We use Qwen3.5 \citep{qwen3} as our base language model, a decoder-only transformer trained on a large corpus of code and natural language. We experiment with three model scales: 4B, 9B, and 27B parameters. All models use the standard Qwen3.5 tokenizer (vocabulary size 151{,}936 tokens) with byte-pair encoding. The 4B model has 32 layers, 32 attention heads, and a hidden dimension of 3{,}584; the 9B and 27B models scale proportionally in depth and width. We chose Qwen3.5 for its strong base performance on code-generation benchmarks and its efficient architecture, which allows us to train at multiple scales within our compute budget.

\paragraph{Training configuration.}
We fine-tune using Low-Rank Adaptation (LoRA) \citep{hu2022lora} with rank 32, targeting all attention projection matrices (\texttt{q\_proj}, \texttt{k\_proj}, \texttt{v\_proj}, \texttt{o\_proj}) and feed-forward layers. LoRA reduces trainable parameters by $\sim$99\% compared to full fine-tuning (e.g., $\sim$40M trainable parameters for the 4B model), enabling efficient multi-scale experimentation. We use the AdamW optimizer with a learning rate of $2\times10^{-5}$, linear warmup over 20 steps, and no learning rate decay. Training runs for 240 steps with a batch size of 32 tasks, each task generating a GRPO group of $K{=}8$ rollouts, for an effective batch size of 256 rollouts per step. Training uses temperature 1.0 for sampling during GRPO rollout generation, while evaluation uses greedy decoding (temperature 0.0) to minimize variance.

\paragraph{Reinforcement learning setup.}
We train with Group Relative Policy Optimization (GRPO) \citep{shao2024deepseekmath} on a hosted RL infrastructure \citep{tinker2025} that handles distributed rollout generation, sandbox execution, and gradient synchronization. Each training step proceeds as follows: (i) for each of 32 tasks in the batch, sample $K{=}8$ complete test suites from the current policy; (ii) execute all suites in parallel sandboxes to collect per-suite mutation scores and per-test first-kill counts; (iii) compute group-relative advantages and per-token segment offsets; (iv) update policy parameters under a KL constraint to the reference policy (the base model frozen at initialization). The reference policy prevents reward hacking and maintains output diversity. The KL coefficient is set to 0.1, balancing optimization speed with stability.

\paragraph{Computational resources.}
Training the 4B model takes approximately 18 hours on 8 NVIDIA A100 GPUs (80GB), including sandbox execution time. The 9B and 27B models require 32 and 64 A100 GPUs respectively, with training times of 36 and 72 hours. Sandbox execution (test execution and mutation scoring) accounts for $\sim$60\% of wall-clock time; we parallelize across 128 sandbox workers per training node to maintain throughput. Total training cost for all three scales (including ablation runs and hyper-parameter sensitivity experiments) is approximately 15{,}000 A100-hours. Evaluation on the five benchmarks is comparatively cheap: scoring 2{,}126 ULT tasks at $N{=}5$ takes $\sim$2 hours on 32 sandbox workers.

\paragraph{Baselines.}
We compare three points on the Qwen3.5-4B family:
(i) the untuned \textbf{base} model, which represents zero-shot test generation capability;
(ii) \textbf{+GRPO}, a single-shot GRPO baseline trained only on the suite-level quality reward (correctness plus mutation, with no conciseness bonus and no segment credit), which isolates the contribution of RL from our specific reward design; and
(iii) \textbf{+MIST-RL} \citep{zhu2026mist}, a multi-turn RL method with a per-step marginal-mutation reward that emits one test per rollout and explicitly penalizes redundancy. MIST-RL represents the strongest prior RL approach for mutation-aware test generation. We reimplement MIST-RL on the same Qwen3.5-4B base and training data for fair comparison; reported numbers are from the fewshot variant (one worked example prepended to the prompt) since it outperforms the zero-shot version.
% Our full method (\textbf{+Ockhamareto}) adds the Pareto-gated
% conciseness bonus and token-level segment credit; individual components are
% ablated in \S\ref{sec:ablation}.

\paragraph{Metrics and scoring protocol.}
Each method emits one suite per task, which we score using the per-test
harness. We report the \emph{mutation score} (the primary quality metric),
\emph{statement} and \emph{branch coverage}, \emph{correctness} (the fraction of
tests passing on the reference implementation), the mean number of tests used
($n_{\text{actual}}$), per-test \emph{efficiency}
(mutation score $\div$ $n_{\text{actual}}$), and the valid-suite rate.
For a test
budget $N$, we follow existing work \citep{huang2026benchmark,testdecision2026,jain2024testgeneval} by truncating each suite to its first $N=5$ tests in source order and
taking the \emph{union} of covered lines, covered branches, and killed mutants over those
$N$ tests: a mutant counts if and only if its first killing test is among the first $N$.
Invalid suites receive a score of 0. This first-$N$ aggregation (Figure~\ref{fig:budget})
exposes the redundancy that suite-level metrics can hide.

% \subsection{Hyper-parameters}
% \label{sec:hparams}

% Table~\ref{tab:hparams} lists the default Ockhamareto configuration.

% \begin{table}[h]
% \centering
% \small
% \begin{tabular}{lrl}
% \toprule
% Param & Default & Meaning \\
% \midrule
% $W_{\text{c}}$            & 0.2  & correctness weight \\
% $W_{\text{m}}$            & 0.8  & mutation weight \\
% $W_{\text{P}}$            & 0.15 & Pareto-membership bonus \\
% $W_{\text{N}}$            & 0.15 & conciseness rank bonus \\
% $W_{\text{seg}}$          & 0.5  & segment-credit weight \\
% SEG\_FAIL\_PENALTY        & 0.1  & per-token fail penalty \\
% batch size                & 32   & tasks per step \\
% group size $K$            & 8    & rollouts per task \\
% LoRA rank                 & 32   & adapter rank \\
% learning rate             & 2e-5 & \\
% temperature (train)       & 1.0  & GRPO sampling \\
% temperature (eval)        & 0.0  & greedy decode \\
% \bottomrule
% \end{tabular}
% \caption{Default Ockhamareto hyper-parameters.}
% \label{tab:hparams}
% \end{table}

\subsection{Per-Test Mutation Harness}
\label{sec:infra}

The segment-credit reward (\S\ref{sec:segment}) \emph{requires} per-test
attribution, which stock mutation harnesses do not provide: they report only
suite-level coverage and mutation. We extend a sandboxed harness
(\texttt{pytest}\,+\,\texttt{cosmic-ray} \citep{cosmicray} inside SandboxFusion
\citep{sandboxfusion}) to record, per test: (i) the covered lines (via
\texttt{pytest --cov-context=test}); (ii) \texttt{per\_test\_mut\_killed}, the
number of mutants for which the test was the \emph{first} to fail; and (iii) a
full trinary (mutant, test) kill matrix (killed / passed / not-run). Mutants are
prebuilt once per task and reused across rollouts, keeping per-rollout reward
latency near 1\,s instead of $\sim$110\,s. The same harness and the same
\texttt{per\_test\_mut\_killed} field drive both training and evaluation.

\subsection{Research Questions}
\label{sec:rq}

Our empirical evaluation is guided by five research questions covering end-to-end effectiveness, component contributions, scaling, hyper-parameter sensitivity, and per-function effectiveness--size trade-offs:

\begin{itemize}[leftmargin=1.4em,itemsep=2pt,topsep=2pt]
    \item \textbf{RQ1: Does Ockhamareto achieve better mutation score and suite conciseness than baselines across multiple benchmarks?}

    We compare Ockhamareto against the base model, a vanilla GRPO baseline, and MIST-RL (the strongest prior RL method) on five held-out benchmarks: ULT (in-distribution), HumanEval+, MBPP+, CodeContests, and TestGenEval-Lite. We measure mutation score, statement/branch coverage, suite size ($n_{\text{actual}}$), and per-test efficiency.

    \item \textbf{RQ2: What is the contribution of each component (Pareto gate and segment credit) to the final performance?}

    We ablate the two core components of Ockhamareto: the Pareto-gated conciseness bonus and token-level segment credit. For each ablation, we train a model with one component removed and evaluate its mutation score, suite size, and per-test efficiency on ULT.

    \item \textbf{RQ3: Does the framework scale to larger models, and how do reward shape and model scale interact?}

    We train Ockhamareto at three scales (4B, 9B, 27B parameters) using the same fixed reward configuration and compare mutation score, suite size, and efficiency. This isolates the effect of model capacity from reward design.

    \item \textbf{RQ4: How sensitive is Ockhamareto to its hyper-parameters, specifically the segment-credit weight and Pareto-bonus weight?}

    We vary $W_{\text{seg}}$ (segment-credit weight) and $W_{\text{P}}$ (Pareto-membership bonus) around their default values while holding other settings fixed, and measure the resulting mutation score, suite size, and efficiency trade-offs.

    \item \textbf{RQ5: What do per-function empirical Pareto fronts reveal about how many tests are worth maintaining?}

    We sample multiple suites per policy for a random 100-function ULT subset, construct pooled (mutation, \#tests) Pareto fronts, and analyze knee location, front size, dominance rates, policy contributions, and correlations with static code metrics.
\end{itemize}

\section{Results}
\label{sec:results}

\subsection{RQ1: Does Ockhamareto achieve better mutation score and suite conciseness than baselines?}
\label{sec:rq1}

To answer RQ1, we compare Ockhamareto against three baselines (base model, vanilla GRPO, MIST-RL) on five held-out benchmarks at $N{=}5$.

\subsubsection{RQ1.1: Performance on in-distribution benchmark (ULT)}

Table~\ref{tab:headline} reports the head-to-head comparison at $N{=}5$ across
the five held-out benchmarks. On the in-distribution eval (ULT), Ockhamareto
reaches $49.9\%$ mutation with $2.60$ tests, \emph{strictly Pareto-dominating}
both RL baselines: $+18.6$~pp mutation \emph{and} $44\%$ fewer tests vs
MIST-RL, $+31.0$~pp mutation and $24\%$ fewer tests vs $+$GRPO. Its per-test
efficiency of $19.2\%$ is $3.4\times$ the base model and $2.9\times$ MIST-RL.

The base model achieves only $14.5\%$ mutation with $2.53$ tests, showing that zero-shot generation struggles with both quality and redundancy. Vanilla GRPO (+GRPO) improves to $18.9\%$ mutation but at the cost of larger suites ($3.43$ tests), indicating that suite-level mutation reward alone cannot prevent bloat. MIST-RL achieves $31.3\%$ mutation but emits $4.67$ tests on average, nearly twice Ockhamareto's suite size, resulting in per-test efficiency of only $6.7\%$ compared to Ockhamareto's $19.2\%$.

\subsubsection{RQ1.2: Generalization to out-of-distribution benchmarks}

The same picture holds out of distribution: on HumanEval+, MBPP+, and
CodeContests, Ockhamareto leads mutation \emph{and} statement/branch coverage
while using the smallest suite in every row, and per-test efficiency stays
$1.7$--$2.2\times$ the MIST-RL number. On the repo-level TestGenEval-Lite
benchmark (Django, sympy, scikit-learn, matplotlib), Ockhamareto's compression
survives cleanly (median $3$ tests per suite; MIST-RL emits $13$), and on the
shared usable subset it still edges every metric under the $N{=}5$ budget
cap. Across all five benchmarks the ranking is the same: Ockhamareto wins
mutation, wins statement/branch coverage, and does it with the smallest
suite.

On HumanEval+ (164 tasks), Ockhamareto achieves $81.6\%$ mutation with $3.10$ tests versus MIST-RL's $74.5\%$ with $4.94$ tests ($+7.1$~pp mutation, $37\%$ fewer tests). On MBPP+ (399 tasks), Ockhamareto reaches $67.8\%$ mutation with $3.12$ tests versus MIST-RL's $64.0\%$ with $4.89$ tests. On CodeContests, where mutants are replaced by real buggy human submissions, Ockhamareto achieves $44.6\%$ fault detection with $3.04$ tests versus MIST-RL's $32.4\%$ with $4.90$ tests, a $+12.2$~pp improvement while using $38\%$ fewer tests.

\begin{table*}[t]
\centering
\small
\begin{tabular}{llrrrrrr}
\toprule
Benchmark & Method & Mut. (\%) & Stmt. (\%) & Branch (\%) & Corr. (\%) & $n_{\text{actual}}$ & Eff. (\%) \\
\midrule
\multirow{4}{*}{ULT}
  & Qwen3.5-4B   & 14.5 & 21.6 & 20.5 & 27.6 & \textbf{2.53} & 5.7 \\
  & + GRPO       & 18.9\,{\scriptsize\textcolor{gray}{(+4.4)}} & 31.0\,{\scriptsize\textcolor{gray}{(+9.4)}} & 28.7\,{\scriptsize\textcolor{gray}{(+8.2)}} & 42.0 & 3.43 & 5.5 \\
  & + MIST-RL    & 31.3\,{\scriptsize\textcolor{gray}{(+16.8)}} & 49.1\,{\scriptsize\textcolor{gray}{(+27.5)}} & 45.9\,{\scriptsize\textcolor{gray}{(+25.4)}} & \textbf{72.7} & 4.67 & 6.7 \\
  & + Ockhamareto  & \textbf{49.9}\,{\scriptsize\textcolor{gray}{(+35.4)}} & \textbf{63.1}\,{\scriptsize\textcolor{gray}{(+41.5)}} & \textbf{62.7}\,{\scriptsize\textcolor{gray}{(+42.2)}} & 68.8 & 2.60 & \textbf{19.2} \\
\midrule
\multirow{4}{*}{HumanEval+}
  & Qwen3.5-4B   & 69.1 & 54.1 & 44.2 & 86.2 & 4.70 & 14.7 \\
  & + GRPO       & 69.2\,{\scriptsize\textcolor{gray}{(+0.1)}} & 53.5\,{\scriptsize\textcolor{gray}{(-0.6)}} & 45.2\,{\scriptsize\textcolor{gray}{(+1.0)}} & 86.5 & 4.60 & 15.0 \\
  & + MIST-RL    & 74.5\,{\scriptsize\textcolor{gray}{(+5.4)}} & 58.9\,{\scriptsize\textcolor{gray}{(+4.8)}} & 49.7\,{\scriptsize\textcolor{gray}{(+5.5)}} & \textbf{96.7} & 4.94 & 15.1 \\
  & + Ockhamareto  & \textbf{81.6}\,{\scriptsize\textcolor{gray}{(+12.5)}} & \textbf{59.8}\,{\scriptsize\textcolor{gray}{(+5.7)}} & \textbf{51.3}\,{\scriptsize\textcolor{gray}{(+7.1)}} & 94.5 & \textbf{3.10} & \textbf{26.3} \\
\midrule
\multirow{4}{*}{MBPP+}
  & Qwen3.5-4B   & 58.5 & 40.3 & 20.2 & 78.6 & 4.68 & 12.5 \\
  & + GRPO       & 57.9\,{\scriptsize\textcolor{gray}{(-0.6)}} & 39.3\,{\scriptsize\textcolor{gray}{(-1.0)}} & 18.8\,{\scriptsize\textcolor{gray}{(-1.4)}} & 78.4 & 4.56 & 12.7 \\
  & + MIST-RL    & 64.0\,{\scriptsize\textcolor{gray}{(+5.5)}} & 43.1\,{\scriptsize\textcolor{gray}{(+2.8)}} & 22.6\,{\scriptsize\textcolor{gray}{(+2.4)}} & 93.1 & 4.89 & 13.1 \\
  & + Ockhamareto  & \textbf{67.8}\,{\scriptsize\textcolor{gray}{(+9.3)}} & \textbf{44.1}\,{\scriptsize\textcolor{gray}{(+3.8)}} & \textbf{23.9}\,{\scriptsize\textcolor{gray}{(+3.7)}} & \textbf{93.7} & \textbf{3.12} & \textbf{21.7} \\
\midrule
\multirow{4}{*}{CodeContests}
  & Qwen3.5-4B   & 11.6 & 24.5 & 23.0 & 13.8 & 3.21 & 3.6 \\
  & + GRPO       & 10.0\,{\scriptsize\textcolor{gray}{(-1.6)}} & 25.7\,{\scriptsize\textcolor{gray}{(+1.2)}} & 23.3\,{\scriptsize\textcolor{gray}{(+0.3)}} & 21.2 & 3.10 & 3.2 \\
  & + MIST-RL    & 32.4\,{\scriptsize\textcolor{gray}{(+20.8)}} & 60.6\,{\scriptsize\textcolor{gray}{(+36.1)}} & 59.0\,{\scriptsize\textcolor{gray}{(+36.0)}} & \textbf{64.1} & 4.90 & 6.6 \\
  & + Ockhamareto  & \textbf{44.6}\,{\scriptsize\textcolor{gray}{(+33.0)}} & \textbf{65.8}\,{\scriptsize\textcolor{gray}{(+41.3)}} & \textbf{64.7}\,{\scriptsize\textcolor{gray}{(+41.7)}} & 61.5 & \textbf{3.04} & \textbf{14.7} \\
\midrule
\multirow{4}{*}{TGE-Lite}
  & Qwen3.5-4B   & 16.6 & 18.6 & 29.1 & 46.7 & 5.00 & 3.3 \\
  & + GRPO       & 13.6\,{\scriptsize\textcolor{gray}{(-3.0)}} & 22.9\,{\scriptsize\textcolor{gray}{(+4.3)}} & 31.6\,{\scriptsize\textcolor{gray}{(+2.5)}} & 76.7 & 5.00 & 2.7 \\
  & + MIST-RL    & 18.4\,{\scriptsize\textcolor{gray}{(+1.8)}} & 17.1\,{\scriptsize\textcolor{gray}{(-1.5)}} & 22.3\,{\scriptsize\textcolor{gray}{(-6.8)}} & \textbf{85.0} & 5.00 & 3.7 \\
  & + Ockhamareto  & \textbf{23.1}\,{\scriptsize\textcolor{gray}{(+6.5)}} & \textbf{23.2}\,{\scriptsize\textcolor{gray}{(+4.6)}} & \textbf{34.8}\,{\scriptsize\textcolor{gray}{(+5.7)}} & 57.6 & \textbf{3.33} & \textbf{6.9} \\
\bottomrule
\end{tabular}
\caption{End-to-end results at $N{=}5$ across five held-out benchmarks. All
methods are Qwen3.5-4B single-shot fewshot; only the training method differs.
CodeContests uses each task's provided incorrect Python submissions as the
fault source (in place of cosmic-ray mutants). TGE-Lite is reported on the
\emph{shared usable set} where every method produces at least one passing
test on the reference module. Ockhamareto leads mutation, statement, and
branch coverage on every benchmark with the smallest suite in every row.}
\label{tab:headline}
\end{table*}

\mybox{\textbf{Answer to RQ1:} Ockhamareto improves fault detection and conciseness simultaneously: it Pareto-dominates the RL baselines on ULT and retains the same qualitative advantage across all four out-of-distribution benchmarks. The first-$N$ curves further show that its gains are concentrated in the earliest tests.}

\begin{figure*}[t]
\centering
\includegraphics[width=\textwidth]{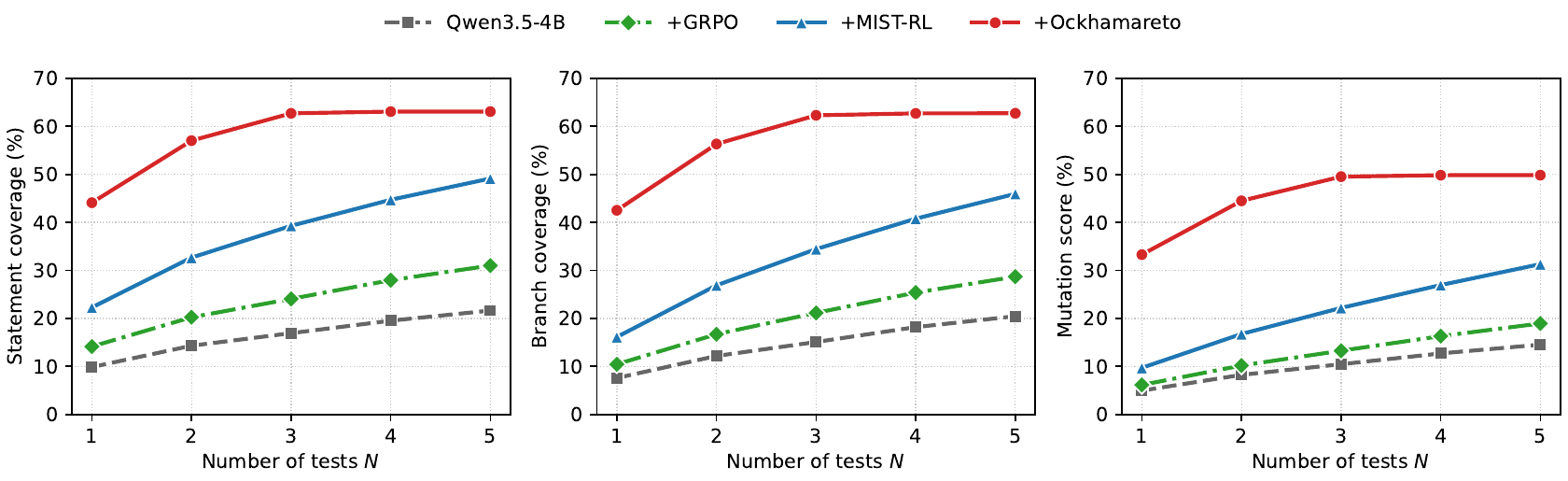}
\caption{First-$N$ union statement coverage, branch coverage, and mutation
score on \textbf{ULT} ($N\!\in\!\{1,\dots,5\}$). Ockhamareto saturates by
$N{=}3$ and dominates all three metrics at every budget; its first test alone
($33.3\%$ mutation) already exceeds MIST-RL's fifth ($31.3\%$).}
\label{fig:budget}
\end{figure*}

\paragraph{Behavior across the test budget (Figure~\ref{fig:budget}).}
Two findings drive the headline. First, \textbf{Ockhamareto saturates by $N{=}3$}:
it captures $99\%$ of its $N{=}5$ mutation in three tests, whereas both
baselines climb steadily and are still below Ockhamareto's $N{=}1$ score at their
own $N{=}5$. Second, \textbf{Ockhamareto's first test alone} ($33.3\%$
mutation at $N{=}1$) already exceeds MIST-RL's fifth ($31.3\%$), a
$5\times$ reduction in suite size for equivalent mutation, the direct effect
of segment credit concentrating reward on the highest-yield test.

\paragraph{Ockhamareto emits shorter suites to begin with.}
The $n_{\text{actual}}$ column in Table~\ref{tab:headline} counts tests inside
the first-$N{=}5$ window, but a sharper picture emerges from the \emph{raw}
suites the policies emit: base and both RL baselines write $12$--$16$ tests per
task on average (median $11$--$12$, $p90$ up to $26$), whereas Ockhamareto writes
only $3.02$ tests per task (median $3$, $p90$ $3$, max $6$).
The baselines' first-$5$ metrics therefore discard $\sim$60\% of the tests
they produced, while Ockhamareto's numbers are essentially unaffected by
truncation. Two mechanisms drive this compression jointly:
segment credit gives the model per-test feedback, so it stops writing tests
that add no mutation; the Pareto bonus then rewards suites whose \emph{total}
size is on the (mutation, $-$\#tests) frontier, so a small policy shift that
drops a marginal test is directly reinforced. Removing either component
recovers the baselines' bloated behavior, quantified in
\S\ref{sec:rq2} and \S\ref{sec:rq4}.

\subsection{RQ2: What is the contribution of each component?}
\label{sec:rq2}

To answer RQ2, Table~\ref{tab:ablation} removes each of Ockhamareto's two core
components: the Pareto-gated conciseness bonus and token-level segment
credit, one at a time on ULT (Qwen3.5-4B, $N{=}5$), while holding the remaining
configuration fixed.
Removing the conciseness bonus ($W_{\text{P}}{=}W_{\text{N}}{=}0$) drops mutation
by $9.9$~pp ($49.9\%\!\to\!40.0\%$) and per-test efficiency by $6.7$~pp
($19.2\%\!\to\!12.5\%$); the model still writes reasonable suites because
segment credit continues to concentrate reward on high-yield tests, but the
Pareto pressure that keeps the suite \emph{small} is gone, and $n_{\text{actual}}$
climbs from $2.60$ to $3.21$. Removing segment credit ($W_{\text{seg}}{=}0$)
shows the opposite failure: mutation falls by $8.7$~pp
($49.9\%\!\to\!41.2\%$) and $n_{\text{actual}}$ grows from $2.60$ to $3.07$,
dragging efficiency down to $13.4\%$. Without per-test credit assignment the
policy reverts to ``more tests catch more mutants,'' recovering some
suite-level mutation only by spending extra tests. The two components thus
guard different failure modes: the Pareto bonus caps suite size, and segment
credit ensures each test carries its own weight. Their weights ($W_{\text{seg}}$
and $W_{\text{P}}$) are varied individually in \S\ref{sec:rq4}.

\begin{table*}[t]
\centering
\small
\begin{tabular}{lrrrrrr}
\toprule
Variant & Mut. (\%) & Stmt. (\%) & Branch (\%) & Corr. (\%) & $n_{\text{actual}}$ & Eff. (\%) \\
\midrule
\textbf{Ockhamareto}                                  & \textbf{49.9} & \textbf{63.1} & \textbf{62.7} & 68.8 & 2.60 & \textbf{19.2} \\
\;$w/o$ segment credit ($W_{\text{seg}}{=}0$)              & 41.2 & 57.8 & 54.2 & 77.1 & 3.07 & 13.4 \\
\;$w/o$ conciseness bonus ($W_{\text{P}}{=}W_{\text{N}}{=}0$) & 40.0 & 56.7 & 55.6 & 67.2 & 3.21 & 12.5 \\
\bottomrule
\end{tabular}
\caption{Component ablation on ULT, Qwen3.5-4B at $N{=}5$. Each row
removes one of the two Ockhamareto components while holding the rest fixed:
the token-level segment credit or the Pareto-gated conciseness bonus.
``Eff.'' is mutants killed per test (Mutation $\div$ $n_{\text{actual}}$).}
\label{tab:ablation}
\end{table*}

\mybox{\textbf{Answer to RQ2:} Both mechanisms are necessary and complementary: Pareto gating primarily controls whole-suite conciseness, while segment credit improves the fault-detection value assigned to individual tests. Removing either substantially reduces mutation efficiency.}

\subsection{RQ3: Does the framework scale to larger models?}
\label{sec:rq3}

To answer RQ3, we train Ockhamareto at three scales (4B, 9B, 27B parameters) using the same fixed reward configuration and compare against the corresponding untuned base models.

Table~\ref{tab:scale-main} pairs each backbone with its untuned
counterpart.
Two effects show up cleanly. First, applying the same fixed Ockhamareto configuration
delivers a $+30$--$35$~pp mutation lift at \emph{every} scale
($+35.4$ at 4B, $+34.5$ at 9B, $+30.5$ at 27B), so the framework's advantage
does not wash out as the backbone grows. Second, raw scale on the base does
help (4B$\to$27B lifts base mutation from $14.5\%$ to $30.5\%$), but the
same $6.75\times$ parameter increase inside the framework only adds $+11.1$~pp
($49.9\!\to\!61.0\%$), less than a third of what the framework itself contributes. The
sharpest way to see this: \emph{Ockhamareto 4B ($49.9\%$) beats base 27B
($30.5\%$) by $+19.4$~pp}, while using $\approx 7\times$ fewer parameters
and $28\%$ fewer tests per suite. 
% Reward shape is the load-bearing lever;
% scale is an additional dial that composes with it but does not replace it.

\begin{table*}[t]
\centering
\small
\begin{tabular}{llrrrrrrr}
\toprule
Scale & Model & Mut. (\%) & Stmt. (\%) & Branch (\%) & Corr. (\%) & $n_{\text{actual}}$ & Eff. (\%) & $\Delta$mut \\
\midrule
\multirow{2}{*}{4B}  & Base       & 14.5 & 21.6 & 20.5 & 27.6 & 2.53 & 5.7  & --- \\
                     & \textbf{Ockhamareto} & \textbf{49.9} & \textbf{63.1} & \textbf{62.7} & \textbf{68.8} & 2.60 & \textbf{19.2} & \textbf{+35.4} \\
\midrule
\multirow{2}{*}{9B}  & Base       & 19.8 & 29.0 & 27.1 & 38.3 & 3.22 & 6.2  & --- \\
                     & \textbf{Ockhamareto} & \textbf{54.3} & \textbf{66.2} & \textbf{65.5} & \textbf{64.2} & 2.71 & \textbf{20.0} & \textbf{+34.5} \\
\midrule
\multirow{2}{*}{27B} & Base       & 30.5 & 39.9 & 38.6 & 54.6 & 3.58 & 8.5  & --- \\
                     & \textbf{Ockhamareto} & \textbf{61.0} & \textbf{70.5} & \textbf{69.8} & \textbf{60.1} & 2.82 & \textbf{21.6} & \textbf{+30.5} \\
\bottomrule
\end{tabular}
\caption{Base vs.\ Ockhamareto at three model scales ($N{=}5$): the same
fixed configuration delivers a $+30$--$35$~pp mutation lift at every scale,
with suite size compressed ($n_{\text{actual}}\!<\!3$) while base suites
grow; base-27B remains $+19.4$~pp behind Ockhamareto-4B.}
\label{tab:scale-main}
\end{table*}

\mybox{\textbf{Answer to RQ3:} Ockhamareto's benefit persists from 4B to 27B parameters, with a $+30$--$35$~pp mutation lift over the corresponding base model at every scale. Model scale and reward design are complementary, but reward design provides the larger gain in these experiments.}

\subsection{RQ4: How sensitive is Ockhamareto to its hyper-parameters?}
\label{sec:rq4}

To answer RQ4, we vary the segment-credit weight $W_{\text{seg}}$ and the
Pareto-membership bonus $W_{\text{P}}$ around their defaults, changing one
weight at a time while holding the remaining configuration fixed.
Table~\ref{tab:hp-sensitivity} reports the trade-off between the effectiveness and size.

\begin{table*}[t]
\centering
\small
\begin{tabular}{lrrrrrr}
\toprule
 & Mut. (\%) & Stmt. (\%) & Branch (\%) & Corr. (\%) & $n_{\text{actual}}$ & Eff. (\%) \\
\midrule
\multicolumn{7}{l}{\emph{Segment-credit weight $W_{\text{seg}}$ ($W_{\text{P}}{=}W_{\text{N}}{=}0.15$ fixed)}} \\
$0.00$          & 41.2 & 57.8 & 54.2 & 77.1 & 3.07 & 13.4 \\
$0.25$          & 45.4 & 64.2 & 63.3 & 74.6 & 2.88 & 15.8 \\
$0.50$ \,(def.) & \textbf{49.9} & \textbf{63.1} & \textbf{62.7} & 68.8 & 2.60 & \textbf{19.2} \\
$0.75$          & 43.1 & 54.7 & 55.0 & 62.5 & 2.18 & 19.8 \\
\midrule
\multicolumn{7}{l}{\emph{Conciseness-bonus weight $W_{\text{P}}{=}W_{\text{N}}$ ($W_{\text{seg}}{=}0.5$ fixed)}} \\
$0.00$          & 40.0 & 56.7 & 55.6 & 67.2 & 3.21 & 12.5 \\
$0.15$ \,(def.) & \textbf{49.9} & \textbf{63.1} & \textbf{62.7} & 68.8 & 2.60 & 19.2 \\
$0.30$          & 47.2 & 61.5 & 60.9 & 67.4 & 2.31 & \textbf{20.4} \\
\bottomrule
\end{tabular}
\caption{Hyper-parameter sensitivity (Qwen3.5-4B, $N{=}5$). Each panel
varies one weight with the other held at its default; the default rows repeat the
headline model. The $W_{\text{seg}}{=}0$ and $W_{\text{P}}{=}0$ rows match
the corresponding ablation rows of Table~\ref{tab:ablation} (the
$W_{\text{P}}$ analysis ties $W_{\text{N}}$ to the same value throughout).}
\label{tab:hp-sensitivity}
\end{table*}

\paragraph{Segment-credit weight $W_{\text{seg}}$ (Table~\ref{tab:hp-sensitivity}, top).}
$W_{\text{seg}}$ controls how much of the reward is delivered as per-token
segment offsets rather than as a single scalar per rollout. Setting
$W_{\text{seg}}{=}0$ recovers pure suite-level Pareto GRPO (this is the
$-$\,segment credit row of Table~\ref{tab:ablation}). We report best-checkpoint
numbers for each $W_{\text{seg}}$ point (step selected by peak mutation at
$N{=}5$). All non-default points lag the default: turning segment credit
off entirely ($W_{\text{seg}}{=}0$) drops mutation to $41.2\%$ at
$n{=}3.07$, the policy loses per-test resolution and pays for mutation with
extra tests; $W_{\text{seg}}{=}0.25$ hits $45.4\%$ mutation at slightly larger
suites ($n{=}2.88$); and $W_{\text{seg}}{=}0.75$ hits only $43.1\%$ with
smaller suites ($n{=}2.18$) because over-weighting the per-token offsets
destabilizes the policy (valid-suite rate drops to $73\%$ vs $85\%$ at the
default). $W_{\text{seg}}{=}0.5$ sits at the knee for mutation, while
$W_{\text{seg}}{=}0.75$ has marginally higher per-test efficiency
($19.8$ vs $19.2$) at the cost of the mutation drop.

\paragraph{Conciseness-bonus weight $W_{\text{P}}$ (Table~\ref{tab:hp-sensitivity}, bottom).}
$W_{\text{P}}$ is the flat bonus awarded to every Pareto-non-dominated rollout,
and we tie $W_{\text{N}}$ (the conciseness-rank bonus) to it throughout.
Setting $W_{\text{P}}{=}W_{\text{N}}{=}0$ turns the conciseness bonus off (this
is the $-$\,conciseness-bonus row of Table~\ref{tab:ablation}) and yields the
lowest mutation and largest suites ($40.0\%$, $n{=}3.21$). Doubling the
default to $W_{\text{P}}{=}W_{\text{N}}{=}0.30$ compresses suites further
($n{=}2.60\!\to\!2.31$) but shaves $2.7$~pp of mutation ($49.9\%\!\to\!47.2\%$)
because the frontier bonus starts to drown out the suite-level mutation
signal; per-test efficiency, however, rises to $20.4\%$, marginally above the
default. The default $W_{\text{P}}{=}W_{\text{N}}{=}0.15$ sits at the raw
mutation peak, while $0.30$ trades a small mutation drop for the best
efficiency among the settings tested.

\mybox{\textbf{Answer to RQ4:} The default weights maximize mutation among the tested settings. Moderate changes preserve the overall benefit, while stronger segment credit or conciseness pressure shifts the balance toward smaller suites and, beyond the default, lower mutation.}

\subsection{RQ5: How Many Tests Should You Maintain? The Empirical Pareto Front}
\label{sec:front}
\begin{figure}[t]
\centering
\includegraphics[width=0.5\columnwidth]{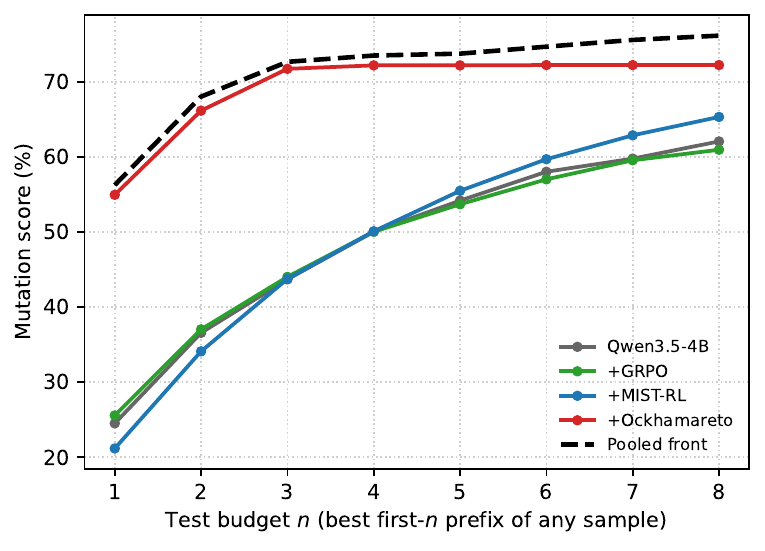}
\caption{The (mutation, \#tests) Pareto trade-off on a $100$-task random
ULT subset: $K{=}50$ suites per policy sampled at $T{=}0.7$; each curve
shows the best mutation achievable with at most $n$ tests, averaged over
tasks. The dashed empirical Pareto front pools all policies' suites and
takes the per-task best; Ockhamareto tracks it within $3$--$5$~pp and
supplies $60.8\%$ of front points.}
\label{fig:front}
\end{figure}

The first-$N$ analysis in \S\ref{sec:rq1} fixes a test budget and asks how much
quality each policy delivers. RQ5 inverts the question: \emph{for a given
function, how many tests are worth maintaining at all?} For every task in a
random 100-task ULT subset, we sample $K{=}50$ suites per policy at temperature
$0.7$ and score each suite as a point in (mutation, \#tests) space.
Figure~\ref{fig:front} reports, for each policy, the best mutation among suites
of at most $n$ tests, averaged over tasks; the dashed empirical Pareto front
pools all four policies' suites and takes the per-task best.

Table~\ref{tab:front-metrics} answers RQ5 with distribution statistics
over the $100$ per-task fronts. The knee sits at a median (and mode) of
$3$ tests, ranging from $1$ to $14$: the typical function repays about
three tests, but individual functions vary by an order of magnitude
(exemplars are shown in \cref{sec:frontgallery}).
The front itself is
small, with a median of $2$ non-dominated points, so the engineer's real
decision space is a handful of discrete choices rather than a continuum.
Most striking is the cost of ignoring the front: of $\approx 165$ valid
sampled suites per task, a median of $97\%$ are \emph{dominated}, with a
median dominated-to-front ratio of $47{:}1$. A suite chosen without front
extraction is therefore almost surely suboptimal, which is the Ockham's
razor principle operationalized: as few tests as possible, but no fewer.

Two further findings concern \emph{who} supplies the front. First,
Ockhamareto's curve tracks the pooled front to within $3$--$5$~pp at every
budget, while MIST-RL trails it by $20$~pp, and Ockhamareto supplies
$60.8\%$ of all per-task pooled-front points, four times the share of any
baseline. Second, the aggregate front has its knee at three tests
($50.0\%$ mutation, about $94\%$ of its $n{=}5$ value). The conclusion is
robust to truncation: allowing every first-$k$ prefix of every sampled
suite to compete, a rule maximally generous to the baselines' large
suites, Ockhamareto still contributes $57\%$ of pooled-front points.

\begin{table}[t]
\centering
\small
\begin{tabular}{lrrrrr}
\toprule
Front metric & Min & Max & Mean & Med. & Mode \\
\midrule
Knee location (\#tests)      & 1    & 14   & 3.6   & 3    & 3 \\
Front size (\#points)        & 1    & 5    & 2.2   & 2    & 2 \\
Dominated suites             & 30   & 198  & 159.6 & 168  & 166 \\
Dominated\,:\,front ratio    & 3.4  & 198  & 66.9  & 47.0 & -- \\
\bottomrule
\end{tabular}
\caption{Distribution of per-function Pareto-front metrics over the
$100$-task study ($\approx 165$ valid sampled suites per task). The median
function has a knee at $3$ tests and a front of only $2$ points, and $97\%$
of sampled suites are dominated: without front extraction, a randomly
chosen suite is almost surely suboptimal.}
\label{tab:front-metrics}
\end{table}

\subsubsection{Per-Function Front Gallery}
\label{sec:frontgallery}

\begin{figure*}
\centering
\includegraphics[width=0.9\textwidth]{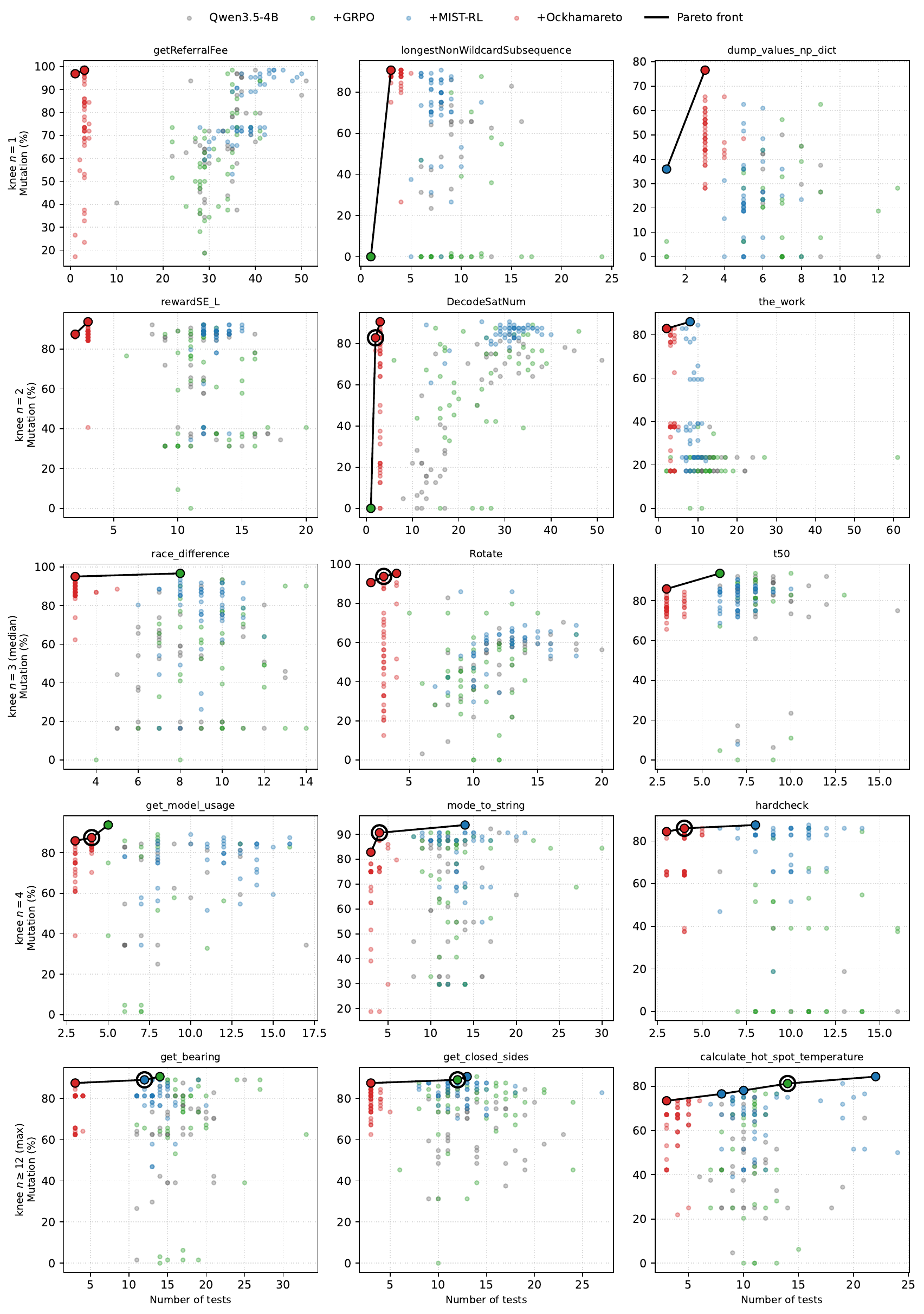}
\caption{Fifteen per-function (mutation, \#tests) sample clouds and pooled
Pareto fronts, stratified by knee location (rows: knee at $1$, $2$, $3$
(the median), $4$, and the maximum end $n\!\ge\!12$; within each row, the
three functions with the highest best-achievable mutation among fronts
with at least two points). Knee points are circled; front vertices are
colored by the supplying policy. The low-budget region of every front
belongs to Ockhamareto; baselines appear only on the diminishing-returns
tail.}
\label{fig:exemplar-fronts}
\end{figure*}

\begin{figure*}
\centering
\includegraphics[width=0.9\textwidth]{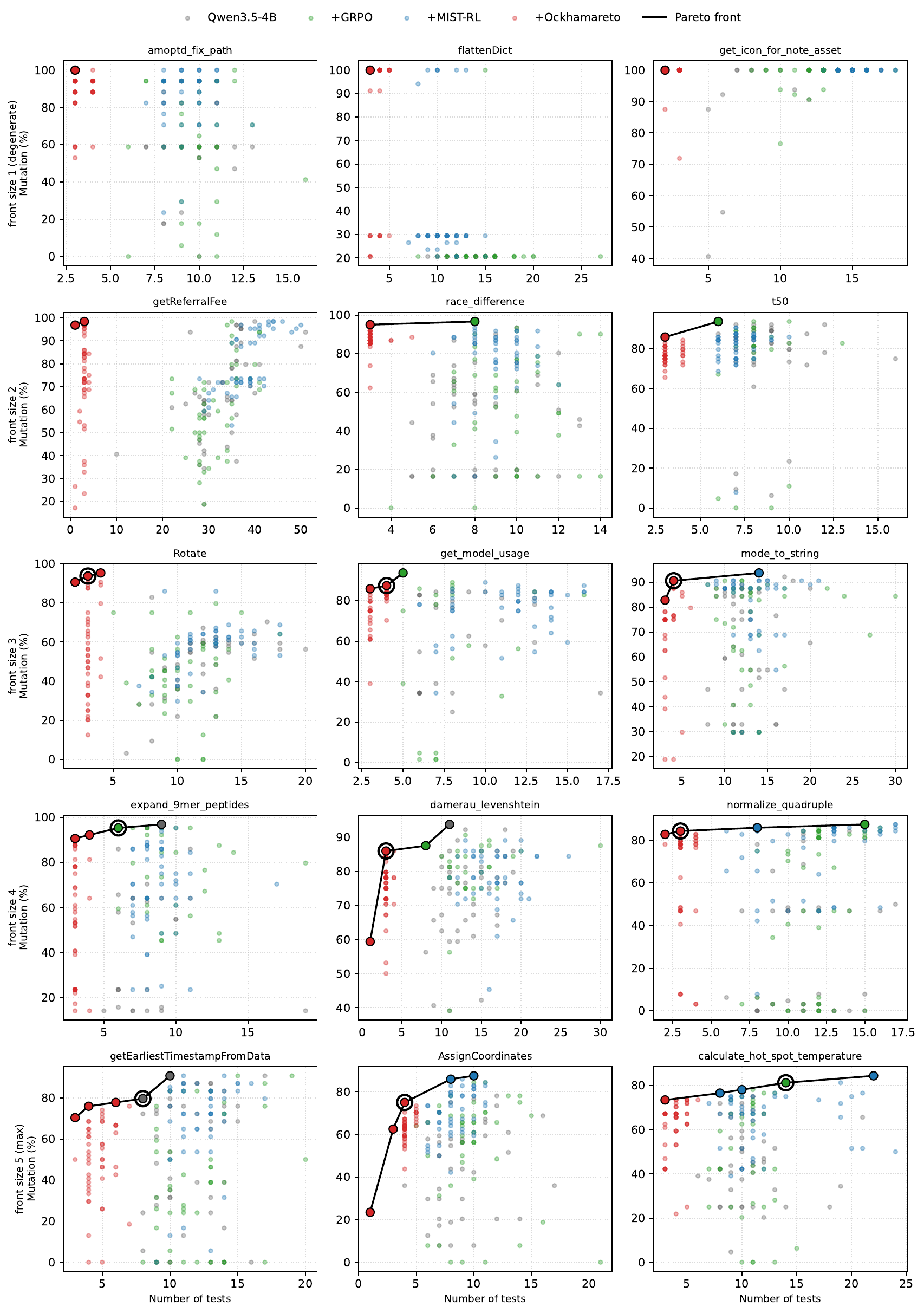}
\caption{
    Fifteen per-function fronts stratified by \emph{front size}
    (rows: $1$ point, the degenerate no-trade-off case, through $5$ points,
    the maximum observed; within each row, the three functions with the
    highest best-achievable mutation). Knee points are circled where defined;
    front vertices are colored by the supplying policy. Degenerate fronts are
    single Ockhamareto points at $100\%$ mutation; larger fronts offer the
    engineer a menu of discrete operating points.
}
\label{fig:size-fronts}
\end{figure*}

The front metrics of Table~\ref{tab:front-metrics} average over functions,
but individual functions differ widely in how many tests they repay.
Figure~\ref{fig:exemplar-fronts} shows fifteen per-function fronts
selected by a stated rule, with no cherry-picking: functions are
stratified by knee location into five rows spanning the observed range,
knees at $1$, $2$, $3$ (the median and mode), $4$, and the maximum end
$n\!\ge\!12$ (exactly three such functions exist, with knees at $12$,
$12$, and $14$); within each stratum, the three functions with the
highest best-achievable mutation among fronts with at least two points
are shown (single-point degenerate fronts, $30\%$ of tasks, draw no curve
here and appear in the front-size view of Figure~\ref{fig:size-fronts}
instead). The rows
illustrate the full spectrum uncovered by RQ5. Functions with a knee at
one or two tests, such as \texttt{DecodeSatNum}, reach over $80\%$
mutation with two tests while every baseline needs $30$ or more to
approach the same level. Typical functions knee at three or four tests.
Maximum-end functions such as \texttt{calculate\_hot\_spot\_temperature}
genuinely repay ten or more tests, and their fronts keep climbing to
twenty tests where the typical function has long saturated. Front
vertices are colored by the policy that supplies them: across all fifteen
panels the low-budget region of the front, including the knee or its entry
point, is supplied by Ockhamareto, while baseline suites appear only on
the diminishing-returns tail. Notably, neither knee position nor front
size correlates with static size or complexity metrics: over the $100$
functions, Kendall's $\tau_b$, which is robust to the heavy ties these
small rank ranges induce, lies between $-0.02$ and $0.08$ against lines
of code and cyclomatic complexity (all $p{>}0.3$; Spearman's $\rho$
agrees), even though the two static metrics correlate strongly with each
other ($\tau_b{=}0.36$, $p{<}0.001$), ruling out a measurement artifact.
The only significant association is internal to the front itself: larger
fronts knee later ($\tau_b{=}0.39$, $p{<}0.001$), a modest but
consistent link between the two dimensions of the trade-off space. How
many tests a function repays is therefore revealed by the empirical
front, not by any static metric we tested, and an engineer can use the
front directly to size the testing opportunity space of each unit and
choose a defensible operating point.

A complementary view stratifies by \emph{front size}, the number of
distinct choices the trade-off space offers (Figure~%
\ref{fig:size-fronts}; observed distribution $1{:}30$, $2{:}34$,
$3{:}24$, $4{:}8$, $5{:}4$). The degenerate row is the extreme of
Ockham's razor: there is literally no trade-off to navigate, and in all
three panels the single optimal point is an Ockhamareto suite reaching
$100\%$ mutation with two or three tests. Each subsequent row adds one
choice, and by the five-point row, for example
\texttt{AssignCoordinates}, the engineer faces a genuine menu, from one
test at $23\%$ mutation to ten tests at $88\%$, with the knee circled
mid-front. Read together, the two grids answer complementary questions:
Figure~\ref{fig:exemplar-fronts} shows \emph{where} the best trade-off
sits, and Figure~\ref{fig:size-fronts} shows \emph{how much choice}
surrounds it.

\mybox{\textbf{Answer to RQ5:} Appropriate suite size is strongly function-specific: the median knee is three tests, but observed knees range from one to fourteen and are not reliably predicted by simple static code metrics. Empirical front extraction is therefore useful for exposing a small set of defensible effectiveness--size operating points, and Ockhamareto supplies most of the pooled-front solutions.}

\section{Discussion}
\label{sec:discussion}

\paragraph{Interpreting the Pareto-gate effect.}
The component ablation suggests that the Pareto gate changes more than output
length. Removing the conciseness bonus increases average suite size from
$2.60$ to $3.21$ tests while mutation falls from $49.9\%$ to $40.0\%$. Thus,
without explicit whole-suite pressure, the policy does not recover the lost
quality simply by generating more tests. This pattern is consistent with the
gate favoring rollouts that occupy a better effectiveness--size position
relative to their peers, rather than rewarding shortness in isolation.

\paragraph{Interpreting segment credit.}
Removing segment credit likewise lowers mutation to $41.2\%$ and increases
suite size to $3.07$ tests. Combined with the first-$N$ result, Ockhamareto's
first test alone reaches $33.3\%$ mutation, this supports the intended
interpretation that fine-grained credit encourages the policy to concentrate
fault-detection value in earlier, higher-yield tests. The evidence therefore
points to complementary roles: suite-level Pareto pressure shapes which suites
are worth generating, while segment credit shapes which tests within those
suites carry the learning signal.

\paragraph{Signal-health diagnostics.}
\label{sec:health}
Because segment credit depends on mapping unit-test blocks to tokens via
offset mapping, Table~\ref{tab:health} reports the fraction of rollouts
for which the mapping succeeds (\texttt{seg\_active}) over the training
run; on mismatch we fall back to scalar GRPO.

\begin{table}[t]
\centering
\small
\begin{tabular}{lr}
\toprule
Diagnostic & Value \\
\midrule
\texttt{seg\_active} rate (token map OK) & 67.8\% \\
mean segments / suite & 3.49 \\
\texttt{sandbox\_fail} rate & 0.15\% \\
\bottomrule
\end{tabular}
\caption{Segment-credit signal health over the training run; rollouts
where re-tokenization does not round-trip ($\sim$32\%) train under
scalar GRPO.}
\label{tab:health}
\end{table}

\paragraph{Deployment implications.}
At any maintainable test budget, Ockhamareto delivers more fault detection than
the alternatives, and the gap is largest at the small budgets that dominate CI
and review cost. Because suites saturate by three tests, most of the value
survives aggressive truncation, making the framework attractive precisely
where budgets are tight.

\section{Threats to Validity}
\label{sec:threats}

\subsection{Internal Validity}

Internal validity concerns potential confounding factors within our experimental setup that could influence observed outcomes. A primary threat pertains to the reproducibility and determinism of the generated test cases. The stochastic nature of some decoding strategies could lead to variability in results, making it difficult to attribute performance differences solely to the reward design. To mitigate this, we employ greedy decoding (temperature 0.0) for all evaluation runs, ensuring that generated outputs are deterministic and reflect the model's learned policy rather than sampling artifacts. During training, we use temperature 1.0 for GRPO rollout sampling, which is standard practice for exploration, but we report all main results from greedy evaluation checkpoints.

Another threat is the reliability of our test execution and mutation-scoring environment. Inconsistencies in the sandbox environment, such as differing library versions or non-deterministic test execution, could lead to spurious failures or inaccurate mutation scores. We address this by executing all test cases within a standardized Docker container with pinned Python interpreter and library versions, and we verify that mutation scoring is deterministic by re-running a random subset of tasks and confirming identical scores.

The token-level segment credit mechanism relies on the tokenizer's offset mapping to align source spans with token boundaries. A mismatch between sampled token IDs and re-tokenized IDs causes the mechanism to fall back to scalar GRPO. Our signal-health diagnostics (Table~\ref{tab:health}) show that offset mapping succeeds for 67.8\% of rollouts; the remaining 32\% train under scalar GRPO, which could dilute the segment-credit signal. However, ablation results (Table~\ref{tab:ablation}) show that even with this partial application, segment credit contributes substantially to final performance.

\subsection{External Validity}

External validity relates to the generalizability of our findings beyond the specific experimental context. One threat is the representativeness of our benchmarks. Although we evaluate on five distinct benchmarks spanning function-level tasks (ULT, HumanEval+, MBPP+), competitive programming (CodeContests), and repo-level tests (TestGenEval-Lite), all benchmarks are Python-only and focus on unit testing. Our results may not generalize to other programming languages (Java, C++, JavaScript) where syntax, testing frameworks, and mutation operators differ. Similarly, our focus on unit-test generation limits generalizability to integration testing or system-level testing, where cross-module dependencies and stateful interactions are the primary challenge.

Another threat is our reliance on a single base model family (Qwen3.5). While we ablate across three scales (4B, 9B, 27B), all share the same architecture and pre-training data distribution. The observed benefits of Pareto-gated conciseness and segment credit may not transfer to models with different tokenization schemes, context lengths, or reasoning capabilities. However, the mechanisms are designed to be model-agnostic: Pareto gating is a group-relative reward filter, and segment credit is a structural decomposition of trajectory-level rewards. Both should apply to any autoregressive language model trained with policy gradient methods.

Finally, our training data (PLT, $\sim$10{,}500 tasks) is drawn from The Stack v2 and decontaminated against evaluation sets, but it remains possible that similar functions or testing patterns appear in the base model's pre-training corpus. This could inflate absolute performance numbers, though it would affect all methods equally since they share the same base model. The rank order of methods and the relative gains from our reward components should remain valid.

\subsection{Construct Validity}

Construct validity examines whether our evaluation metrics and experimental design accurately measure the concepts they purport to assess. A central threat is our reliance on mutation score as the primary quality metric. While mutation testing is widely regarded as a stronger proxy for real-fault detection than coverage \citep{just2014mutants,papadakis2019mutation}, it is not perfect: some mutants are equivalent (semantically identical to the original), some are trivial (caught by any test that executes the line), and mutation score does not directly measure a suite's ability to catch real bugs in production. Nonetheless, mutation score provides a quantitative, reproducible, and well-validated adequacy criterion that correlates strongly with fault detection in empirical studies.

Another threat is our use of per-test first-kill attribution to drive segment credit. A test that is the \emph{first} to kill a mutant receives full credit for that kill, even if later tests would also have detected it. This biases the signal toward tests that appear earlier in the suite. However, this bias is intentional: we \emph{want} the policy to front-load its most discriminating tests, since in practice the first few tests are the ones most likely to be run frequently (e.g., in pre-commit hooks or incremental CI). The first-$N$ evaluation protocol (Figure~\ref{fig:budget}) directly measures this front-loading behavior.

The Pareto-gated conciseness bonus treats mutation score and test count as the two dimensions of a bi-objective optimization problem, with equal weight given to both axes in dominance comparisons. This design choice encodes a specific preference: a suite that catches one more mutant with one more test is considered non-comparable (neither dominates the other) rather than strictly better. An alternative design might weight the axes differently or use a scalarized objective. We chose the unweighted Pareto formulation because it avoids committing to a specific mutation-versus-size exchange rate, which is likely to vary across deployment contexts. The hyper-parameter sensitivity analysis (Table~\ref{tab:hp-sensitivity}) shows that the design is robust to moderate changes in $W_{\text{P}}$ and $W_{\text{seg}}$.

\section{Related Work}
\label{sec:related}

\paragraph{Classical test generation.}
Before LLMs, automated test generation was dominated by search-based and
random approaches: EvoSuite evolves suites to maximize coverage
\citep{fraser2011evosuite}, Pynguin brings the same idea to Python
\citep{lukasczyk2022pynguin}, and feedback-directed random testing builds
suites incrementally \citep{pacheco2007randoop}. These tools optimize coverage
and tend to emit large suites; minimization is handled, if at all, as a
\emph{post hoc} reduction step. We instead make conciseness a first-class
\emph{training} objective, so the policy never learns to produce the redundancy
in the first place.

\paragraph{Mutation testing.}
Mutation testing measures a suite's fault-detection power by seeding small
faults and checking whether tests catch them \citep{jia2011mutation,
papadakis2019mutation}. It is widely regarded as a stronger adequacy criterion
than coverage. We use a sandboxed \texttt{cosmic-ray} \citep{cosmicray} mutant
pool as both the reward signal and the evaluation metric, and, crucially,
attribute kills at \emph{per-test} granularity to drive token-level credit.

\paragraph{Fine-grained credit and process rewards.}
A central limitation of trajectory-level RL is coarse credit assignment: a
scalar return is shared uniformly across all tokens. In reasoning, \emph{process
reward models} that score intermediate steps rather than only final answers
improve learning \citep{lightman2023verify}. Our segment-credit mechanism is a
domain-specific analogue: rather than a learned step verifier, it derives an
\emph{exact, executable} per-step signal, namely each test's marginal mutation
kills, and maps it back to that test's tokens via the tokenizer's offset
mapping. The decomposition is natural and lossless: a suite is literally a
sequence of independently scorable test functions. To our knowledge
this is the first use of per-test execution outcomes as a token-level RL signal
for test generation.

\paragraph{LLMs for test generation.}
Prompt-based methods elicit tests from pre-trained models directly from code and
documentation: by intention-then-refinement \citep{yuan2024chattester}, by
per-path coverage-guided prompting \citep{ryan2024symprompt}, by hybridizing with
search \citep{lemieux2023codamosa,chen2021codex}, or by mining usage examples
\citep{schafer2024testpilot}. They optimize primarily for coverage and
compilation/pass rates, and empirical studies report that the resulting suites
are often redundant and smell-laden
\citep{lin2024empirical,siddiq2024empirical}. We depart from them by treating
\emph{mutation score} as the quality target and \emph{test count} as an explicit
cost, optimizing the trade-off rather than coverage alone.

\paragraph{RL for code and test generation.}
RL from execution feedback is now standard for code LLMs
\citep{ouyang2022instructgpt,deepseekr1}, and GRPO \citep{shao2024deepseekmath}
replaces PPO's value network \citep{schulman2017ppo} with group-relative
advantages, a natural fit for verifiable-reward settings. For test generation,
recent RL methods reward execution outcomes: TestCTRL optimizes a learned
coverage reward \citep{testctrl2025}, TestDecision rewards each test's marginal
coverage gain in a submodular decision process \citep{testdecision2026}, and
MIST-RL rewards each test's marginal mutation kills \citep{zhu2026mist}. Our two closest neighbors, TestDecision and MIST-RL, both obtain per-test
credit through \emph{multi-turn} generation, emitting and evaluating one test
at a time. This makes attribution straightforward but requires sequential LLM
calls and optimizes the value of each addition rather than directly pricing
the effectiveness--size position of the complete suite. Ockhamareto instead
combines \textbf{(i)} single-shot generation, \textbf{(ii)} a whole-suite
Pareto conciseness signal, and \textbf{(iii)} intra-trajectory token-level
credit. The last mechanism can be viewed as an exact, execution-derived
analogue of process rewards \citep{lightman2023verify}.

\section{Conclusion}
\label{sec:conclusion}

Prior RL methods for unit-test generation treat bug-catching power and suite
size as competing objectives. We showed they need not compete: a single-shot
GRPO framework that combines a Pareto-gated conciseness bonus with token-level
segment credit shifts the entire (mutation, $-$\#tests) frontier, so
Ockhamareto-4B \emph{strictly Pareto-dominates} both RL baselines on a held-out
benchmark (more bugs \emph{and} fewer tests), saturates mutation within
three tests, and front-loads bug-catching power into its first. The framework
scales cleanly to 9B and 27B (adding $+30$--$35$~pp mutation at every scale),
and Ockhamareto-4B even outperforms base-27B. For this task, principled reward shaping is a
stronger lever than raw model scale, and the two compose.

\small
\section{Generative AI Disclosure}
Claude and GPT-5.6 Sol were used to assist with
code development (Claude),
proofreading (Claude and GPT-5.6 Sol),
improving the clarity of author-written text (Claude and GPT-5.6 Sol),
and
identifying potentially relevant literature (Claude and GPT-5.6 Sol).
All AI-assisted content included in this paper was reviewed by,
and remains the responsibility of, at least one human author.

\noindent
{\bf Note:} At the time of writing, ACM policy does not require disclosure for the use of generative AI solely for editing and refining author-written text.
We nevertheless disclose all such use here, together with other uses for which disclosure may be required,
in the interests of full transparency and to future-proof the scholarly record against possible changes in publisher policies and community standards.
% \balance
\bibliographystyle{acmart-primary/ACM-Reference-Format}
\bibliography{acmart-primary/acmart}

@article{shao2024deepseekmath,
  title={Deepseekmath: Pushing the limits of mathematical reasoning in open language models},
  author={Shao, Zhihong and Wang, Peiyi and Zhu, Qihao and Xu, Runxin and Song, Junxiao and Bi, Xiao and Zhang, Haowei and Zhang, Mingchuan and Li, YK and Wu, Yang and others},
  journal={arXiv preprint arXiv:2402.03300},
  year={2024}
}

@article{schulman2017ppo,
  title={Proximal policy optimization algorithms},
  author={Schulman, John and Wolski, Filip and Dhariwal, Prafulla and Radford, Alec and Klimov, Oleg},
  journal={arXiv preprint arXiv:1707.06347},
  year={2017}
}

@article{hu2022lora,
  title={Lora: Low-rank adaptation of large language models.},
  author={Hu, Edward J and Shen, Yelong and Wallis, Phillip and Allen-Zhu, Zeyuan and Li, Yuanzhi and Wang, Shean and Wang, Liang and Chen, Weizhu and others},
  journal={Iclr},
  volume={1},
  number={2},
  pages={3},
  year={2022}
}

@article{ouyang2022instructgpt,
  title={Training language models to follow instructions with human feedback},
  author={Ouyang, Long and Wu, Jeffrey and Jiang, Xu and Almeida, Diogo and Wainwright, Carroll and Mishkin, Pamela and Zhang, Chong and Agarwal, Sandhini and Slama, Katarina and Ray, Alex and others},
  journal={Advances in neural information processing systems},
  volume={35},
  pages={27730--27744},
  year={2022}
}

@article{qwen3,
  title={Qwen3 technical report},
  author={Yang, An and Li, Anfeng and Yang, Baosong and Zhang, Beichen and Hui, Binyuan and Zheng, Bo and Yu, Bowen and Gao, Chang and Huang, Chengen and Lv, Chenxu and others},
  journal={arXiv preprint arXiv:2505.09388},
  year={2025}
}

@article{jia2011mutation,
  title={An analysis and survey of the development of mutation testing},
  author={Jia, Yue and Harman, Mark},
  journal={IEEE transactions on software engineering},
  volume={37},
  number={5},
  pages={649--678},
  year={2010},
  publisher={IEEE}
}

@incollection{papadakis2019mutation,
  title={Mutation testing advances: an analysis and survey},
  author={Papadakis, Mike and Kintis, Marinos and Zhang, Jie and Jia, Yue and Le Traon, Yves and Harman, Mark},
  booktitle={Advances in computers},
  volume={112},
  pages={275--378},
  year={2019},
  publisher={Elsevier}
}

@inproceedings{lemieux2023codamosa,
  title={Codamosa: Escaping coverage plateaus in test generation with pre-trained large language models},
  author={Lemieux, Caroline and Inala, Jeevana Priya and Lahiri, Shuvendu K and Sen, Siddhartha},
  booktitle={2023 IEEE/ACM 45th International Conference on Software Engineering (ICSE)},
  pages={919--931},
  year={2023},
  organization={IEEE}
}

@article{schafer2024testpilot,
  title={An empirical evaluation of using large language models for automated unit test generation},
  author={Sch{\"a}fer, Max and Nadi, Sarah and Eghbali, Aryaz and Tip, Frank},
  journal={IEEE Transactions on Software Engineering},
  volume={50},
  number={1},
  pages={85--105},
  year={2023},
  publisher={IEEE}
}

@article{zhu2026mist,
  title={MIST-RL: Mutation-based Incremental Suite Testing via Reinforcement Learning},
  author={Zhu, Sicheng and Wang, Jiajun and Ai, Jiawei and Li, Xin},
  journal={arXiv preprint arXiv:2603.01409},
  year={2026}
}

@article{testdecision2026,
  title={TestDecision: Sequential Test Suite Generation via Greedy Optimization and Reinforcement Learning},
  author={Wang, Guoqing and Yang, Chengran and Zhou, Xiaoxuan and Sun, Zeyu and Wang, Bo and Lo, David and Hao, Dan},
  journal={arXiv preprint arXiv:2604.01799},
  year={2026}
}

@article{testctrl2025,
  title={Automated Unit Test Generation via Chain-of-Thought Prompt and Reinforcement Learning from Coverage Feedback},
  author={Zhang, Junwei and Hu, Xing and Xia, Xin and Cheung, Shing-Chi and Li, Shanping},
  journal={ACM Transactions on Software Engineering and Methodology},
  volume={35},
  number={4},
  pages={1--30},
  year={2026},
  publisher={ACM New York, NY}
}

@article{yuan2024chattester,
  title={No more manual tests? evaluating and improving chatgpt for unit test generation},
  author={Yuan, Zhiqiang and Lou, Yiling and Liu, Mingwei and Ding, Shiji and Wang, Kaixin and Chen, Yixuan and Peng, Xin},
  journal={arXiv preprint arXiv:2305.04207},
  year={2023}
}

@article{ryan2024symprompt,
  title={Code-aware prompting: A study of coverage-guided test generation in regression setting using llm},
  author={Ryan, Gabriel and Jain, Siddhartha and Shang, Mingyue and Wang, Shiqi and Ma, Xiaofei and Ramanathan, Murali Krishna and Ray, Baishakhi},
  journal={Proceedings of the ACM on Software Engineering},
  volume={1},
  number={FSE},
  pages={951--971},
  year={2024},
  publisher={ACM New York, NY, USA}
}

@inproceedings{just2014mutants,
  title={Are mutants a valid substitute for real faults in software testing?},
  author={Just, Ren{\'e} and Jalali, Darioush and Inozemtseva, Laura and Ernst, Michael D and Holmes, Reid and Fraser, Gordon},
  booktitle={Proceedings of the 22nd ACM SIGSOFT international symposium on foundations of software engineering},
  pages={654--665},
  year={2014}
}

@article{lin2024empirical,
  title={An empirical study of unit test generation with large language models},
  author={Yang, Lin and Yang, Chen and Gao, Shutao and Wang, Weijing and Wang, Bo and Zhu, Qihao and Chu, Xiao and Zhou, Jianyi and Liang, Guangtai and Wang, Qianxiang and others},
  journal={arXiv preprint arXiv:2406.18181},
  year={2024}
}

@inproceedings{siddiq2024empirical,
  title={Using large language models to generate junit tests: An empirical study},
  author={Siddiq, Mohammed Latif and Da Silva Santos, Joanna Cecilia and Tanvir, Ridwanul Hasan and Ulfat, Noshin and Al Rifat, Fahmid and Carvalho Lopes, Vin{\'\i}cius},
  booktitle={Proceedings of the 28th international conference on evaluation and assessment in software engineering},
  pages={313--322},
  year={2024}
}

@misc{cosmicray,
  title={Cosmic Ray: Mutation Testing for {Python}},
  author={{Sixty North}},
  year={2024},
  howpublished={\url{https://github.com/sixty-north/cosmic-ray}}
}

@misc{sandboxfusion,
  title={{SandboxFusion}: A Sandbox for Code Execution and Evaluation},
  author={{ByteDance}},
  year={2024},
  howpublished={\url{https://github.com/bytedance/SandboxFusion}}
}

@misc{tinker2025,
  title={Tinker: A Hosted Service for Reinforcement Learning Fine-Tuning},
  author={{Thinking Machines Lab}},
  year={2025},
  howpublished={\url{https://thinkingmachines.ai}}
}

@article{chen2021codex,
  title={Evaluating large language models trained on code},
  author={Chen, Mark and Tworek, Jerry and Jun, Heewoo and Yuan, Qiming and Pinto, Henrique Ponde De Oliveira and Kaplan, Jared and Edwards, Harri and Burda, Yuri and Joseph, Nicholas and Brockman, Greg and others},
  journal={arXiv preprint arXiv:2107.03374},
  year={2021}
}

@article{deepseekr1,
  title={Deepseek-r1: Incentivizing reasoning capability in llms via reinforcement learning},
  author={Guo, Daya and Yang, Dejian and Zhang, Haowei and Song, Junxiao and Wang, Peiyi and Zhu, Qihao and Xu, Runxin and Zhang, Ruoyu and Ma, Shirong and Bi, Xiao and others},
  journal={arXiv preprint arXiv:2501.12948},
  year={2025}
}

@inproceedings{fraser2011evosuite,
  title={Evosuite: automatic test suite generation for object-oriented software},
  author={Fraser, Gordon and Arcuri, Andrea},
  booktitle={Proceedings of the 19th ACM SIGSOFT symposium and the 13th European conference on Foundations of software engineering},
  pages={416--419},
  year={2011}
}

@inproceedings{lukasczyk2022pynguin,
  title={Pynguin: Automated unit test generation for python},
  author={Lukasczyk, Stephan and Fraser, Gordon},
  booktitle={Proceedings of the ACM/IEEE 44th International Conference on Software Engineering: Companion Proceedings},
  pages={168--172},
  year={2022}
}

@inproceedings{pacheco2007randoop,
  title={Feedback-directed random test generation},
  author={Pacheco, Carlos and Lahiri, Shuvendu K and Ernst, Michael D and Ball, Thomas},
  booktitle={29th International Conference on Software Engineering (ICSE'07)},
  pages={75--84},
  year={2007},
  organization={IEEE}
}

@inproceedings{lightman2023verify,
  title={Let's verify step by step},
  author={Lightman, Hunter and Kosaraju, Vineet and Burda, Yuri and Edwards, Harrison and Baker, Bowen and Lee, Teddy and Leike, Jan and Schulman, John and Sutskever, Ilya and Cobbe, Karl},
  booktitle={International Conference on Learning Representations},
  volume={2024},
  pages={39578--39601},
  year={2024}
}

@inproceedings{
liu2023evalplus,
title={Is Your Code Generated by Chat{GPT} Really Correct? Rigorous Evaluation of Large Language Models for Code Generation},
author={Jiawei Liu and Chunqiu Steven Xia and Yuyao Wang and LINGMING ZHANG},
booktitle={Thirty-seventh Conference on Neural Information Processing Systems},
year={2023},
url={https://openreview.net/forum?id=1qvx610Cu7}
}

@article{
  li2022codecontests,
  author = {Yujia Li  and David Choi  and Junyoung Chung  and Nate Kushman  and Julian Schrittwieser  and R{\'e}mi Leblond  and Tom Eccles  and James Keeling  and Felix Gimeno  and Agustin Dal Lago  and Thomas Hubert  and Peter Choy  and Cyprien de Masson d’Autume  and Igor Babuschkin  and Xinyun Chen  and Po-Sen Huang  and Johannes Welbl  and Sven Gowal  and Alexey Cherepanov  and James Molloy  and Daniel J. Mankowitz  and Esme Sutherland Robson  and Pushmeet Kohli  and Nando de Freitas  and Koray Kavukcuoglu  and Oriol Vinyals },
  title = {Competition-level code generation with AlphaCode},
  journal = {Science},
  volume = {378},
  number = {6624},
  pages = {1092-1097},
  year = {2022},
  doi = {10.1126/science.abq1158},
  URL = {https://www.science.org/doi/abs/10.1126/science.abq1158},
  eprint = {https://www.science.org/doi/pdf/10.1126/science.abq1158}}

@inproceedings{
jain2024testgeneval,
title={TestGenEval: A Real World Unit Test Generation and Test Completion Benchmark},
author={Kush Jain and Gabriel Synnaeve and Baptiste Roziere},
booktitle={The Thirteenth International Conference on Learning Representations},
year={2025},
url={https://openreview.net/forum?id=7o6SG5gVev}
}

@inproceedings{
jimenez2024swebench,
title={{SWE}-bench: Can Language Models Resolve Real-world Github Issues?},
author={Carlos E Jimenez and John Yang and Alexander Wettig and Shunyu Yao and Kexin Pei and Ofir Press and Karthik R Narasimhan},
booktitle={The Twelfth International Conference on Learning Representations},
year={2024},
url={https://openreview.net/forum?id=VTF8yNQM66}
}

@article{huang2026benchmark,
author = {Huang, Dong and Zhang, Jie M. and Harman, Mark and Zhang, Qianru and Du, Mingzhe and Ng, See-Kiong},
title = {Benchmarking LLMs for Unit Test Generation from Real-World Functions},
year = {2026},
publisher = {Association for Computing Machinery},
address = {New York, NY, USA},
issn = {1049-331X},
url = {https://doi.org/10.1145/3805043},
doi = {10.1145/3805043},
note = {Just Accepted},
journal = {ACM Trans. Softw. Eng. Methodol.},
month = mar
}

@InProceedings{turing:checking,
  author =       "Alan M. Turing",
  title =        "Checking a Large Routine",
  booktitle =    "Report of a Conference on High Speed Automatic
                 Calculating Machines",
  publisher =    "University Mathematical Laboratory",
  address =      "Cambridge, England",
  month =        jun,
  year =         "1949",
  pages =        "67--69",
  entered =      "7 October 1992 from C. B. Jones Survey",
}

@book{Ockham1323,
  author    = {{William of Ockham}},
  title     = {Ockham's Theory of Terms: Part I of the Summa Logicae},
  note      = {Originally written c. 1323. Translated and introduced
               by Michael J. Loux},
  publisher = {University of Notre Dame Press},
  address   = {Notre Dame, Indiana},
  year      = {1974}
}

@inproceedings{symh:issta07,
author = "Shin Yoo and Mark Harman",
title = "Pareto Efficient Multi-Objective Test Case Selection",
booktitle = "International Symposium on Software Testing and Analysis ({ISSTA'07})",
publisher = acm,
address = "London, United Kingdom",
month = "July", 
year = 2007,
pages = "140 -- 150"  
}

@book{Pareto1906,
  author     = {Vilfredo Pareto},
  title      = {Manual of Political Economy},
  note       = {Originally published in Italian as
                \emph{Manuale di Economia Politica} in 1906},
  translator = {Ann S. Schwier},
  publisher  = {Augustus M. Kelley},
  address    = {New York},
  year       = {1971}
}

@misc{dijksra:aphorism,
author =	"Edsger W. Dijkstra",
title = "Structured programming",
url = "http://www.cs.utexas.edu/users/EWD/ewd02xx/EWD268.PDF",
year = 1969,
note = "circulated privately"
}

\end{document}